%% file: bare_jrnl.tex
\documentclass[journal]{IEEEtran}
\usepackage{cite}
\usepackage{amsmath,amssymb}
\usepackage{cleveref}
\usepackage{algorithm}
\usepackage{algorithmic}
\usepackage[caption=false,font=footnotesize]{subfig}
\usepackage{booktabs}
\usepackage{graphicx}
\usepackage{float}
\usepackage{soul}   
\usepackage{xcolor} 
\usepackage[normalem]{ulem}  

\ifCLASSINFOpdf
\else
\fi

\begin{document}

\title{Iterative Receiver Processing at Relays in PNC-Enabled Multi-Hop Underwater Acoustic Networks}

\author{Gewei~Zhang,~\IEEEmembership{Member,~IEEE,}
        Deqing~Wang,~\IEEEmembership{Member,~IEEE,}
        Lizhao~You,~\IEEEmembership{Member,~IEEE,}
        Xiangming~Cai,~\IEEEmembership{Member,~IEEE,}
        and Liqun~Fu,~\IEEEmembership{Senior~Member,~IEEE}}

\maketitle

\begin{abstract}
Physical-layer network coding (PNC) can increase end-to-end throughput in bi-directional multi-hop underwater acoustic (UWA) networks. However, multipath delay spread and Doppler-induced inter-carrier interference (ICI) in UWA channels can degrade the reliability of PNC transmission in a three-node relay configuration. More critically, error accumulation across multiple relay nodes leads to a pronounced increase in the end-to-end bit error rate (BER) in multi-hop networks. To address this issue, we develop an iterative detection and decoding processing strategy for relay nodes within a PNC-enabled multi-hop UWA network based on orthogonal frequency division multiplexing (OFDM) modulation. The proposed design integrates three key algorithms: (i) an adaptive channel-aware factor graph detection algorithm that is suited for time-varying UWA channels; (ii) a parity-check-constrained soft-information refinement algorithm that improves the accuracy of the information feedback from the decoder to the detector; and (iii) a linear minimum mean square error (LMMSE) detection algorithm based on a superimposed model, which offers low computational complexity as an alternative scheme. Extensive simulation results demonstrate that the adaptive detection algorithm achieves BERs on the order of $10^{-5}$ at a relative velocity of 1.5 m/s UWA channel and a signal-to-noise (SNR) of 8~dB. Both lake experiments and sea trials in the Taiwan Strait confirm that the proposed iterative receiver algorithms outperform baseline schemes in terms of BER performance under practical UWA channel conditions, showing their robustness and applicability in real multi-hop deployments.

\end{abstract}


\IEEEpeerreviewmaketitle

\section{Introduction}\label{sec:introduction}

\IEEEPARstart{U}{nderwater} acoustic (UWA) communication and networking technology play an important role in environmental monitoring, resource exploration, and data collection \cite{li2025nature}. Among various marine applications, the demand for long-distance monitoring and data transmission has been steadily increasing \cite{yang2024Review}. Current research focuses on bidirectional multi-hop networks, such as autonomous underwater vehicle (AUV)-assisted seabed surveys, buoy-based environmental monitoring\cite{guo2025ASMAC}, and underwater pipeline monitoring\cite{jawhar2018architecture}. In multi-hop scenarios, end-to-end throughput is the key performance metric. However, in traditional half-duplex decode-and-forward relaying, collision avoidance and scheduling mechanisms require repeated handshakes, resulting in additional propagation delays. As the number of hops increases, the accumulated delay reduces the end-to-end throughput.

\begin{figure}[!t]
      \centering
      \includegraphics[width=0.70\linewidth]{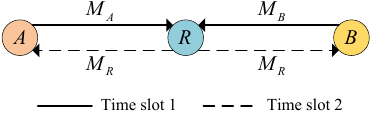}
    \caption{A two-way relay channel network using physical layer network coding.}
    \label{fig:PNC}
\end{figure}

\begin{figure}[!t]
    \centering
    \includegraphics[width=0.85\linewidth]{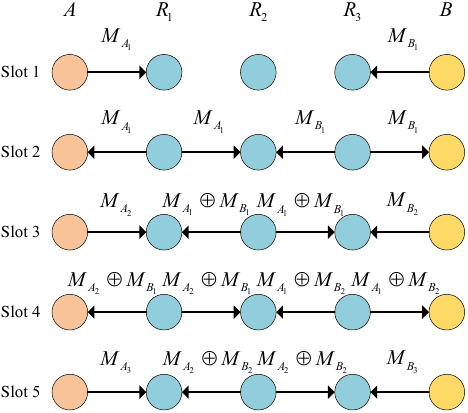} 
    \caption{A bi-directional multi-hop network using PNC with 3 relay nodes.}
    \label{fig:pnc_mh}
\end{figure}

Physical layer network coding (PNC) has demonstrated potential in improving end-to-end throughput and has seen substantial advancements in wireless relay networks \cite{zhang2006hot,you2017rpnc,cpp2025pnc}. PNC treats collision interference as a coding opportunity, making it effective for multi-hop networks that require complex message exchanges~\cite{zhang2015pipnc, lin2015throughput, naves2018framework}. As depicted in Fig.~\ref{fig:PNC}, in a two-way relay channel (TWRC) network, the end nodes $A$ and $B$ exchange their respective message packets, $M_A$ and $M_B$, via a relay node $R$. Messages from both end nodes are transmitted simultaneously to the relay node in the multiple access phase. By applying a network coding operation to the superimposed signals, the relay node obtains a combined message $M_R = M_A \oplus M_B$ and broadcasts it to both end nodes. By retaining its original message, each end node can recover the message of the other node from the network-coded symbols, thereby enabling information exchange. Fig.~\ref{fig:pnc_mh} illustrates a multi-hop PNC network consisting of three relay nodes. Node $A$ transmits messages $M_{A_1}, M_{A_2}, \ldots$ to node $B$, while node $B$ sends messages $M_{B_1}, M_{B_2}, \ldots$ to node $A$. During end-to-end transmission, PNC can resolve up to 3 times superposed collision events for a message packet, thereby improving throughput.

However, the design and deployment of multi-hop UWA PNC face several key challenges. First, traditional PNC operates in a three-node single-relay system, where the impact of propagation errors is limited. In a multi-hop network, errors at any relay can propagate and affect subsequent nodes~\cite{zhang2017design, zhang2017bi}. Second, in multi-hop UWA networks, signal propagation delays are larger than signal processing delays. If the relay node fails to process the received signal, continuous retransmissions will degrade communication efficiency~\cite{wang2016multi, ma2019hybrid, peng2024interference}. Third, most existing multi-hop PNC designs consider static channels and overlook the interference between multiple nodes in dynamic channels. Compared with terrestrial radio channels, UWA channels suffer from multipath delays with a lower sound speed (approximately 1500 m/s) and experience time-varying Doppler shifts, resulting in frequency offsets \cite{qarabaqi2013statistical, van2013propagation, soc2025MSML}. Overall, the most significant point in deploying multi-hop UWA PNC networks is to ensure the reliability of relay reception and processing, which enables low message exchange latency and high throughput.

In current research on UWA TWRC \cite{wang2013iterative, kulhandjian2015cdma}, only \cite{wang2013iterative} explicitly addressed the interference mitigation process. Although orthogonal frequency division multiplexing (OFDM) protects the signal, beyond the length of the cyclic prefix (CP), the transmitted signal experiences inter-carrier interference (ICI) in the frequency domain. UWA channels exhibit a multi-scale multi-lag (MSML) property that renders the frequency-domain channel matrix dense with more off-diagonal elements. Unlike terrestrial wireless communication methods, we cannot eliminate ICI caused by multiple frequency shifts through carrier frequency offset (CFO) compensation \cite{MSML2015}. To address this issue, a channel coding-assisted PNC system based on iterative signal processing has been proposed \cite{wang2013iterative, situ2018ofdm, woltering2017factor, xie2018channel}. The effectiveness and practicability of such iterative architecture has been well demonstrated in point-to-point UWA communication systems \cite{xi2019frequency, li2021enhanced, zheng2025delay}. We adopt the Markov-based factor graph detection presented in \cite{zhou2014ofdm} to mitigate ICI using forward–backward message passing and embed it into the iterative detection and decoding framework, serving as the I-MFGD benchmark. Building upon similar structure, \cite{situ2018ofdm} further reduces computational complexity by localizing belief propagation (BP) updates in an iterative receiver, which is regarded as I-BP benchmark in our comparisons. Both of them model inter‑symbol dependencies via a Markov process and restrict message exchange to adjacent symbols. Unlike the aforementioned methods, our proposed approach employs a factor graph model \cite{woltering2017factor, xie2018channel}, which represents the dependencies between transmitted symbols and their corresponding received symbols through variable and factor nodes. Its advantage lies in its ability to handle large-scale parallel processing while better exploiting \textit{a priori} information.

We further introduce innovative designs to address the unique challenges of relay reception and processing in multi-hop networks.

\textbf{We propose factor-graph-based adaptive channel-aware detection and soft-information refinement.} Although the factor graph framework provides a new perspective for signal detection, it still exhibits limitations. The time-varying nature of UWA channels leads to substantial fluctuations in frequency offsets and interference vary across subcarriers. As a result, relying on a single \emph{ICI depth} is insufficient to cope with the diverse UWA conditions. Since the frequency-domain channel matrix of a UWA link can often be approximated as a banded structure \cite{Zhou2008Nonuniform}, implying that each symbol is influenced by a limited set of adjacent symbols, we adapt the local dependency range based on the channel energy distribution, achieving optimal performance while maintaining low computational complexity.

Recognizing that the performance of UWA communication systems is linked to the gains achieved by channel decoding \cite{Zhao2023channelcode}, we employ LDPC coding. The soft-information from channel decoding is used as \textit{a priori} knowledge for detection, but potential errors in this may degrade the detection performance. Consequently, we refine the soft-information using the parity-check constraints. It enables detection to obtain more reliable \textit{a priori} inputs and improves the reliability of symbol estimation across iterations.

\textbf{We derive a low complexity and superimposed-signal-based solution.} Both the Markov model and the factor graph structure incur considerable computational complexity. It motivates the need for a low-complexity detection alternative even at the cost of performance degradation. Although linear minimum mean-square error (LMMSE) detection was introduced in \cite{bhat2012decoding} for terrestrial wireless channels, it is not integrated into an iterative detection and decoding framework. Deriving a linear solution for multi-hop PNC systems differs from conventional point-to-point communication. Detection needs to exploit the network-coding dependencies embedded in the superimposed signals rather than treating each user independently. By leveraging the \textit{a priori} information provided by channel decoding, we derive an iterative LMMSE detection tailored to the superposition model. The proposed method offers the additional advantage of requiring no channel preconditioning and can be deployed on a per-subcarrier.

\begin{figure*}[t]
    \centering
    \includegraphics[width=\textwidth]{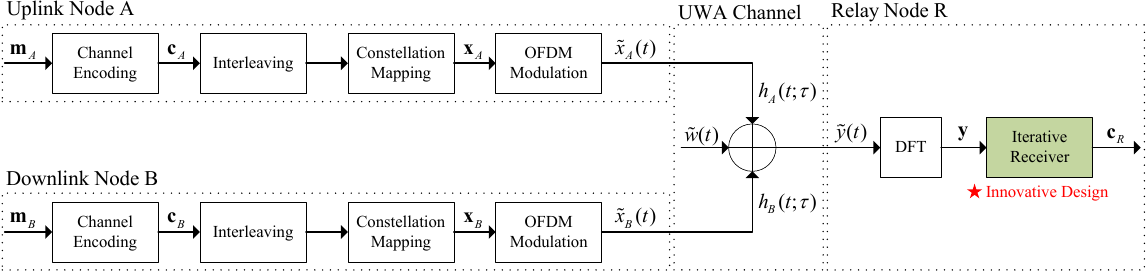} 
    \caption{Illustration of relay reception and processing at the relay in multi-hop UWA PNC networks: the uplink and the downlink node exchange information with each other via the help of the relay node $R$ which processes the superimposed signal at the receiving hydrophone.}
    \label{fig:UWA}
\end{figure*}

Our main contributions are summarized as follows:

\begin{itemize}
\item Within an iterative detection and decoding architecture, we design an adaptive channel-aware factor graph detection (ACA-FGD) algorithm to exploit the unique ICI distribution characteristics of UWA channels. We propose a parity-check-constraint-based soft-information refinement algorithm to enhance the reliability of feedback probabilities across iterations.  

\item We derive a low-complexity and closed-form LMMSE-based signal detection under an iterative detection and decoding framework. It overcomes the limitation of conventional methods that cannot jointly process superimposed signal, while offering simple deployment without requiring any channel preconditioning. 

\item We validate the proposed methods through numerical simulations and real-world field experiments in multi-hop UWA networks. The ACA-FGD algorithm demonstrates robust performance across various UWA channels, achieving an approximately 1~dB BER improvement over conventional methods under multi-hop transmission. In addition, the proposed parity-constraint-based soft information refinement provides up to 0.5~dB BER gain after multiple relay hops. While linear detection exhibits limited effectiveness in multi-hop scenarios, it remains competitive in UWA channels where the energy is highly concentrated along the dominant propagation path.

\end{itemize}

The remainder of this paper is organized as follows. Section~II introduces the MSML model of the UWA channels and derives the physical representation of the received procedure at the relay. Section~III presents the proposed factor-graph-based algorithms and low-complexity linear solution within an iterative detection and decoding framework. In Section~IV, we validate the effectiveness of the proposed algorithms through comprehensive simulations. Section~V presents the experimental results obtained in real-world lake and ocean environments, followed by detailed discussions. Finally, Section~VI concludes the paper.

\textit{Notation:} Bold lowercase and uppercase letters denote vectors and matrices, respectively. For a matrix $\mathbf{A}$, $\mathbf{A}^H$ represents its Hermitian transpose, and $\mathbf{A}^{-1}$ denotes its inverse. $\lVert \mathbf{A} \rVert$ denotes the Euclidean norm, and $\mathbf{A}[m,n]$ refers to the element in the $m$-th row and the $n$-th column of $\mathbf{A}$. $\mathrm{diag}(\mathbf{A})$ denotes the diagonal matrix whose main diagonal consists of the diagonal entries of $\mathbf{A}$. 
Similarly, for a vector $\mathbf{z}$, $\mathrm{diag}(\mathbf{z})$ denotes the diagonal matrix with $\mathbf{z}$ on its main diagonal. $\mathbb{R}$ denotes the set of real numbers and $\Re(\cdot)$ extracts the real part of a complex variable. $j=\sqrt{-1}$ denotes the imaginary unit. $\mathcal{CN}(\mu,\sigma^{2})$ denotes a complex circularly symmetric Gaussian random variable with mean $\mu$ and variance $\sigma^{2}$. $\mathcal{N}(\cdot)$ denotes a set defined by a given constraint and $\sim(\cdot)$ represents a marginalization. $\mathrm{Pr}(\cdot)$ denotes the probability, and $\mathbb{E}[\cdot]$ is written as the expectation of the variable. For two random variable vectors $\mathbf{x}$ and $\mathbf{y}$, $\mathbf{R}_{\mathbf{xy}}$ denotes their cross-covariance, while $\mathbf{R}_{\mathbf{yy}}$ denotes the auto-covariance of $\mathbf{y}$. $\mathbf{I}_{N}$ denotes an identity matrix of size $ N $.

\section{Preliminary and System Model} \label{sec:system model}

We consider an OFDM-based multi-hop UWA PNC system. Similarly to the transmission structure illustrated in Fig.~\ref{fig:pnc_mh}, end-to-end communication in the multi-hop network is achieved through successive relay receiving and forwarding. Since identical signal processing is performed at each relay, the multi-relay signal processing can be reduced to a single relay node meeting message collisions between the uplink and the downlink node. As depicted in Fig.~\ref{fig:UWA}, the information from the uplink node $A$ and downlink node $B$ is transmitted simultaneously to the relay node $R$. Without loss of generality, it is assumed that the both UWA channels associated with the uplink and downlink nodes share identical parameters. The purpose of this section is to establish the underlying system model, which serves as the foundation for the innovative receiver designs presented in Sections~III.

\subsection{Transmitted Signal at the Uplink and Downlink Nodes}

\begin{figure*}[t]
    \centering
    \includegraphics[width=\textwidth]{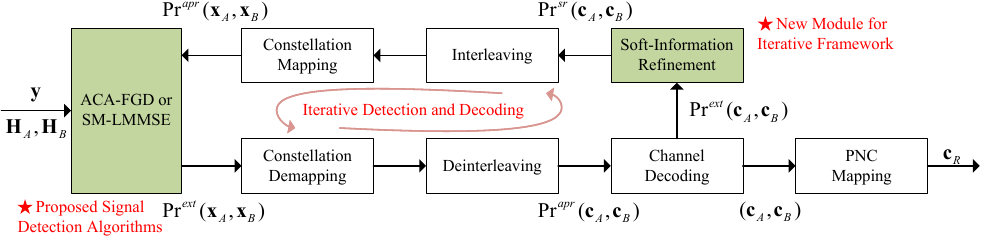}
    \caption{The structure of iterative receiver, which primarily includes three modules: signal detection, channel decoding and soft-information refinement. After handling the superimposed signal, the relay node performs PNC mapping of the estimated $ (\mathbf{c}_A,\mathbf{c}_B) $ to obtain $ \mathbf{c}_R $, which will be subsequently packaged and sent to uplink and downlink nodes.}
    \label{fig:IR}
\end{figure*}

For the uplink or downlink node $ u \in \{A, B\} $, the information bits $ \mathbf{m}_u $ are first encoded with a QC-LDPC code defined over $ \mathrm{GF}(2) $, yielding the codeword $ \mathbf{c}_u $. The encoded sequence is interleaved and mapped to modulated symbols $ \mathbf{x}_u $. Each symbol is drawn from the constellation set $ \mathcal{L}_u = \{ \alpha_1, \alpha_2, \ldots, \alpha_{2^Q} \} $, where $ Q $ denotes the number of bits per symbol. 

Let $ T $ be the duration of the useful OFDM symbol and $ T_{cp} $ be the length of CP. The sampling rate is $ f_s $ and the FFT size is $ N_{F} $. Consequently, the subcarrier spacing is given by $ \Delta f = {f_s / {N_{F}}} = {1/T} $. The modulated results are divided into multiple groups within an OFDM symbol, which contains $ N $ valid data. The frequency of the $ m $-th subcarrier is given by

\begin{equation}
f_m = f_c + \frac{m - N/2}{T}, \quad m \in \{0,1,\ldots,N-1\},
\end{equation}
where $ f_c $ is the carrier frequency. The corresponding passband OFDM symbol for the uplink or downlink node $ u $ can be expressed as

\begin{equation}
\tilde{x}_u(t) = 2 \Re \left\{ \sum_{m=0}^{N-1} \mathbf{x}_u[m] \, e^{j 2\pi f_m t} \right\} ,
\end{equation}
where $ t \in [-T_{cp},T]$.

\subsection{Multi-Scale Multi-Lag UWA Channels}

UWA channels exhibit \emph{doubly spread characteristics} arising from the combined effects of multipath propagation and Doppler spread. The relatively slow propagation speed of sound in water results in pronounced delay spreads, whereas motion between the transmitter and receiver gives rise to Doppler-induced frequency dispersion. To reflect these properties, the MSML model is adopted in \cite{MSML2015}, which captures jointly time-frequency dispersion. 

In the \emph{continuous-time domain}, the channel impulse response (CIR) at the uplink or downlink node $ u $ is described in \cite{wang2013iterative}.

\begin{equation}
h_{u}(t;\tau) = \sum_{p=1}^{N_{P,u}} A_{p,u} \cdot \delta(\tau - (\tau_{p,u} - a_{p,u} t)) ,\label{eq:hu}
\end{equation}
where $ N_{P,u} $ denotes the number of multipath components at node $ u $, and $ A_{p,u} $ represents the amplitude of the $ p $-th path. The parameters $ \tau_{p,u} $ and $ a_{p,u} $ correspond to the initial delay and the Doppler factor, respectively. For analytical simplicity and without loss of generality, the velocities of different paths are assumed identical. In this case, the Doppler factor becomes directly related to the common path velocity, and can be written as

\begin{equation}
a_{p,u} =\frac{\sigma_{u}}{c},
\end{equation}
where $ \sigma_{u} $ denotes the relative velocity between the transmitter and receiver along the direction of propagation, and $ c $ is the speed of sound in water. 

\subsection{Superimposed Received Signal at the Relay Node}

Incorporating the MSML model, the superimposed passband signal observed at the relay can be expressed as

\begin{align}
\tilde{y}(t) = & \sum_{p=1}^{N_{P,A}} A_{p,A} \tilde{x}_{A} \left( (1 + a_{p,A}) t - \tau_{p,A} \right) \nonumber \\
& + \sum_{p=1}^{N_{P,B}} A_{p,B} \tilde{x}_{B} \left( (1 + a_{p,B}) t - \tau_{p,B} \right) + \tilde{w}(t),
\end{align}
where $ \tilde{w}(t) $ denotes additive white Gaussian noise (AWGN). 

It is noted that the signals transmitted from the uplink and downlink nodes can be synchronized at the relay node. The received signal in the time domain is passed through bandpass filtering and OFDM demodulation via the discrete Fourier transform (DFT), yielding the corresponding frequency domain vector. The equivalent subcarrier-wise model can then be represented as

\begin{equation}
\mathbf{y} = \mathbf{H}_{A}\mathbf{x}_A + \mathbf{H}_{B}\mathbf{x}_B + \mathbf{w},
\end{equation}
where $ \mathbf{y} $ denotes the received signal vector in the frequency domain, $ \mathbf{x}_u $ is the transmitted symbol vector, and $ \mathbf{H}_{u} $ represents the equivalent frequency-domain channel matrix between uplink or downlink node $u$ and the relay $R$. The frequency-domain noise vector follows $ \mathbf{w} \sim \mathcal{CN}(\mathbf{0}, \sigma^2 \mathbf{I}_N) $.

In particular, the $ (n,m) $-th entry of $ \mathbf{H}_u $ is obtained through the DFT operation and can be written as

\begin{equation}
\mathbf{H}_{u}[n, m] = \sum_{p=1}^{N_{P,u}} A_{p,u} \rho_{p,u}^{n,m} e^{-j 2\pi f_m \tau_{p,u}},
\end{equation}
where $ \rho_{p,u}^{n,m} $ denotes the coefficient of intercarrier coupling between the $ m $-th transmitted and the $ n $-th received subcarrier induced by ICI. Specifically, the ICI coefficient can be described as

\begin{equation}
\begin{split}
\rho_{p,u}^{n,m} 
&= \mathrm{sinc} \!\left[ \left( a_{p,u} f_m - (f_n - f_m) \right) T \right] \\
&\quad \times e^{j \pi \left( a_{p,u} f_m - (f_n - f_m) \right) T},
\end{split}
\label{eq:rho}
\end{equation}
where $ \mathrm{sinc}(x) = \frac{\sin(\pi x)}{\pi x} $ denotes the normalized sinc function. The quantities $ f_m $ and $ f_n $ correspond to the $ m $-th and $ n $-th subcarrier frequencies, respectively.

\begin{figure*}[!t]
    \centering
    \includegraphics[width=\textwidth]{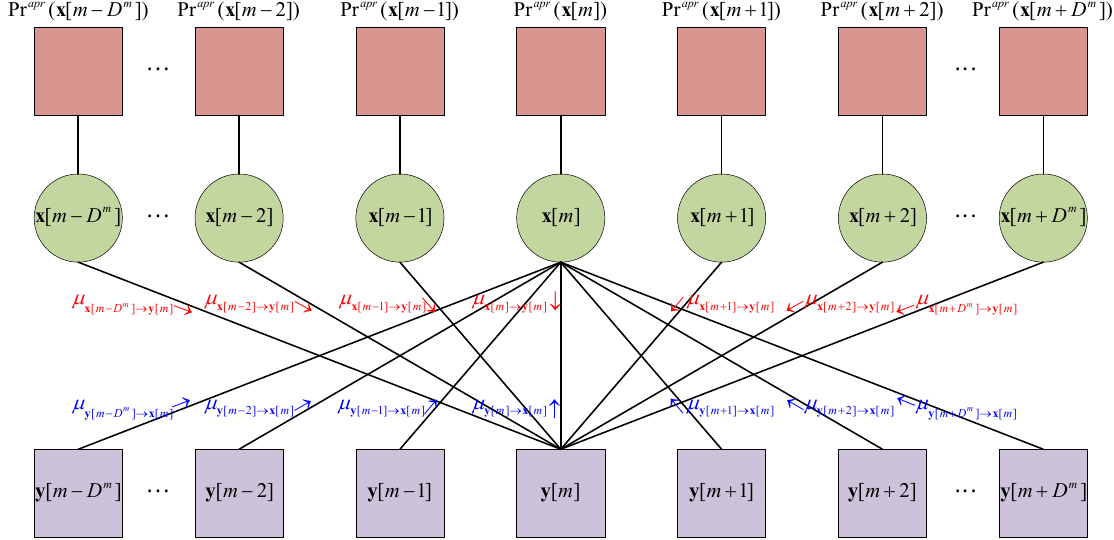}
    \caption{Adaptive channel-aware factor graph signal detection at the $m$-th subcarrier, accounting for mutual interference within a neighborhood of $D^m$ adjacent subcarriers.}
    \label{fig:MP_total}
\end{figure*}

Physically, $ \mathbf{H}_u $ is a full channel-coupling matrix incorporating multipath fading, delay spread, and Doppler frequency spread. The diagonal entries represent the direct subcarrier gains, while the off-diagonal entries quantify ICI caused by channel time variation. For computational tractability, such channels are often regarded as banded matrices. The effective ICI bandwidth of this banded structure is characterized by an \emph{ICI depth} $ D $. It is defined as the number of adjacent subcarriers on either side of a given subcarrier whose mutual coupling with it causes non-negligible interference \cite{wang2013iterative}. All coefficients outside this range are approximated to zero, 

\begin{equation} \label{eq:D}
\mathbf{H}_{u}[n, m] \approx 0, \quad \forall |n - m| > D.
\end{equation}

Accurate determination of $ D $ is essential, as it defines the set of subcarriers that must be considered in effective ICI mitigation. 
This confines the ICI contribution to a limited bandwidth, while ignoring interference beyond $ D $. 

\section{Proposed Iterative Detection and Decoding Receiver at the Relay}\label{sec:iterative framewok}

In this section, we propose iterative receiver design, which constitutes the core innovation of this work and enables more accurate signal detection.

\subsection{Iterative Detection and Decoding Framework with Soft-Information Refinement}

The proposed iterative receiver is developed under the assumption that the equivalent frequency-domain channel matrices $ \mathbf{H}_A $ and $ \mathbf{H}_B $ are well-known at the relay. Under this condition, the receiver architecture is organized into three modules: (i) signal detection, (ii) channel decoding, and (iii) soft-information refinement, as illustrated in Fig.~\ref{fig:IR}. Among these modules, signal detection one plays the central role, processing the superimposed signal received from both uplink and downlink nodes. Using the known channel matrices, signal detection is employed to produce the extrinsic probability distribution $ \mathrm{Pr}^{ext}(\mathbf{x}_A, \mathbf{x}_B) $ of the transmitted symbols.

Unlike conventional iterative receivers, the proposed design incorporates a dedicated soft-information refinement module. The channel decoding first demaps and deinterleaves $ \mathrm{Pr}^{ext}(\mathbf{x}_A, \mathbf{x}_B) $ to obtain the \textit{a priori} probabilities for the coded bits, denoted as $ \mathrm{Pr}^{apr}(\mathbf{c}_A, \mathbf{c}_B) $. In this framework, LDPC codes are employed. After decoding, the \textit{a posteriori} $ \mathrm{Pr}^{ext}(\mathbf{c}_A, \mathbf{c}_B) $ together with parity-check results is forwarded to the refinement module, which updates the extrinsic information $ \mathrm{Pr}^{sr}(\mathbf{c}_A, \mathbf{c}_B) $. After interleaving and mapping, the refined information is fed back to the signal detection module as updated \textit{a priori} probability $ \mathrm{Pr}^{apr}(\mathbf{x}_A, \mathbf{x}_B) $. This iterative exchange among $ \mathrm{Pr}^{ext}(\mathbf{x}_A, \mathbf{x}_B) $, $ \mathrm{Pr}^{apr}(\mathbf{c}_A, \mathbf{c}_B) $, $ \mathrm{Pr}^{ext}(\mathbf{c}_A, \mathbf{c}_B) $, $ \mathrm{Pr}^{sr}(\mathbf{c}_A, \mathbf{c}_B) $, and $ \mathrm{Pr}^{apr}(\mathbf{x}_A, \mathbf{x}_B) $ continues until convergence or a preset maximum iteration.

Finally, the estimated codewords $ ({\mathbf{c}}_A, {\mathbf{c}}_B) $ are mapped at the relay via bitwise XOR to generate $ \mathbf{c}_R $. It is repackaged into an OFDM symbol and broadcast to uplink and downlink nodes.

\subsection{Adaptive Channel-Aware Factor Graph Signal Detection}

To exploit the PNC mapping structure, we construct a joint detection-based factor graph, in which the superimposed signal is represented as a joint variable, expressed as

\begin{equation}
\mathbf{x} = (\mathbf{x}_A, \mathbf{x}_B).
\end{equation}

\begin{figure*}[!t]
\centering
\begin{equation}\label{eq:eta_tD}
D_{u}^{m,*} = 
\underset{D_{u}^{m}}{\arg\min}
\left\{
\frac{
\displaystyle
\sum_{k:\,|k-m|\le D_{u}^{m}}\!\left\lVert\mathbf{H}_u[m,k]\right\lVert^2
+ \sum_{i:\,|i-m|\le D_{u}^{m}}\!\left\lVert\mathbf{H}_u[i,m]\right\lVert^2
- \left\lVert\mathbf{H}_u[m,m]\right\lVert^2
}{
\displaystyle
\sum_{k=0}^{N-1}\left\lVert\mathbf{H}_u[m,k]\right\lVert^2
+ \sum_{i=0}^{N-1}\left\lVert\mathbf{H}_u[i,m]\right\lVert^2
- \left\lVert\mathbf{H}_u[m,m]\right\lVert^2
}
\ge \eta_{u}^{m}
\right\}.
\end{equation}
\hrule
\end{figure*}

Increasing the number of connections between nodes in a factor graph may enhance performance, but excessive connections can introduce too many loops, resulting in performance degradation. Conversely, an overly sparse graph may fail to capture sufficient information for reliable inference. Therefore, accurate detection is fundamentally based on effective channel modeling.

However, during signal transmission, UWA channels exhibit time variation. Although \eqref{eq:D} introduces the notion of a band-limited channel, it does not provide practical guidance on determining the appropriate \emph{ICI depth}, thereby limiting its applicability. Moreover, since Doppler-induced frequency offsets affect the subcarriers within an OFDM symbol in a nonuniform manner, a single global parameter $D$ is generally insufficient to characterize the channel-wide ICI structure. To enable more precise detection, we refine the \emph{ICI depth} on a per-subcarrier basis by examining the lateral and vertical energy distributions around the main diagonal of each frequency-domain channel slice. For the $m$-th subcarrier, \eqref{eq:eta_tD} provides the criterion for depth selection. The optimal depth $D_{u}^{m,*}$ is chosen as the smallest $D_{u}^{m}$ whose energy ratio exceeds the threshold $\eta_{u}^{m}$. When the estimated \emph{ICI depths} for the same subcarrier differ across channels, the maximum value is selected to ensure consistent modeling accuracy.

\begin{equation}\label{eq:Dm}
D^{m} = \max(D_{A}^{m,*}, D_{B}^{m,*}).
\end{equation}

Fig.~\ref{fig:MP_total} highlights the local factor graph associated with the $ m $-th subcarrier, where the factor node $ \mathbf{y}[m] $ is affected by up to $ 2 \times D^m $ adjacent subcarriers, while the variable node $ \mathbf{x}[m] $ contributes effective interference within the same range. Within the iterative detection and decoding workflow, the variable node also incorporates \textit{a priori} information $ \mathrm{Pr}^{apr}(\mathbf{x}[m]) $. In this paper, we illustrate the detection procedure by focusing on the message exchange over the $m$-th subcarrier as a representative example. The detailed steps are described as follows.

\textit{Step 1 Initiation:} The messages exchanged between the variable nodes and the factor nodes, together with the \textit{a priori} information, are initialized to a uniformly distributed vector, i.e., $\mathbf{1}/4^{Q}$.

\textit{Step 2 Depth Selection:} When the superimposed signal and channel information are first fed into the receiver, the optimal depth $D_{u}^{m,*}$ is computed for each subcarrier. The detection depth for the $m$-th subcarrier is chosen as the larger.

\textit{Step 3 Output Messages from Variable Nodes:} For the $m$-th variable node $\mathbf{x}[m]$, we identify the set of its neighboring factor nodes, denoted by $\mathcal{N}(v_m)$. According to the sum–product algorithm (SPA), the message passed from $\mathbf{x}[m]$ to the factor node $\mathbf{y}[m]$ is expressed by

\begin{equation}
\mu_{\mathbf{x}[m] \rightarrow \mathbf{y}[m]} \propto \mathrm{Pr}^{\mathrm{apr}}(\mathbf{x}[m]) \prod_{\substack{n = m - D^{m} \\ n \ne m}}^{m + D^{m}} \mu_{\mathbf{y}[n] \rightarrow \mathbf{x}[m]}.
\label{eq:spaxy}
\end{equation}

\textit{Step 4 Output Messages from Factor Nodes:} Similarly, once the set of variable nodes associated with the factor node $\mathbf{y}[m]$ is identified as $\mathcal{N}(f_m)$, the message passed from $\mathbf{y}[m]$ to $\mathbf{x}[m]$ is given by

\begin{equation}
\mu_{\mathbf{y}[m] \rightarrow \mathbf{x}[m]} = \sum_{\sim (\mathbf{x}[m]) } \left( \mathrm{Pr}_m(\mathbf{x}) \prod_{\substack{n = m - D^{m} \\ n \ne m}}^{m + D^{m}} \mu_{\mathbf{x}[n] \rightarrow \mathbf{y}[m]} \right),
\label{eq:spayx}
\end{equation}
where $ \mathrm{Pr}_m(\mathbf{x}) $ represents a likelihood function in the $ m $-th received subcarrier, written as

\begin{equation}
\begin{aligned}
& \mathrm{Pr}_m(\mathbf{x}) \\ 
& \propto \exp \left( -\frac{1}{\sigma^2} 
\left\lVert 
\mathbf{y}[m] - \sum_{u \in \{A,B\}} \sum_{k=m-D^m}^{m+D^m} \mathbf{H}_u[m,k] \cdot \mathbf{x}_u[k]
\right\rVert^2
\right).
\label{eq:fa}
\end{aligned}
\end{equation}

\begin{algorithm}[!t]
\caption{Proposed ACA-FGD Algorithm}
\label{alg:aca}
\begin{algorithmic}[1]

\STATE \textbf{Initiation:} $ \mu_{\mathbf{x} \rightarrow \mathbf{y}} $, $ \mu_{\mathbf{y} \rightarrow \mathbf{x}} $ and $ \mathrm{Pr}^{apr}(\mathbf{x}) = \mathbf{1}/4^Q $;
\STATE \textbf{Input:} received frequency-domain vector $ \mathbf{y} $, frequency-domain channel matrices $ \mathbf{H}_{A} $ and $ \mathbf{H}_B $, noise power $ \sigma^2 $, \textit{a priori} probability $ \mathrm{Pr}^{apr}(\mathbf{x}_A,\mathbf{x}_B) $;
\STATE {// \textit{Depth Selection} //}
\IF{first detection in the iterative receiver}
    \FORALL{ $ m = 0 $ to $ N-1 $}
        \STATE combine \eqref{eq:eta_tD} to determine $D_{u}^{m,*}$;
        \STATE perform \eqref{eq:Dm} to choose $ D^m $;
    \ENDFOR
\ENDIF
\STATE {// \textit{Output Messages from Variable Nodes} //}
\FOR{$ m = 0 $ to $ N-1 $}
    \STATE find all factor nodes connected to $ \mathbf{x}[m] $: $ \mathcal{N}(f_m) $;
    \FOR{$ n \in \mathcal{N}(f_m) $}
        \STATE using to \eqref{eq:spaxy}, compute $  \mu_{\mathbf{x}[m] \rightarrow \mathbf{y}[n]} $;
    \ENDFOR
\ENDFOR
\STATE {// \textit{Output Messages from Factor Nodes} //}
\FOR{$ m = 0 $ to $ N-1 $}
    \STATE find all variable nodes connected to $ \mathbf{y}[m] $: $ \mathcal{N}(v_m) $;
    \FOR{$ n \in \mathcal{N}(v_m) $}
        \STATE using to \eqref{eq:fa} and \eqref{eq:spayx}, compute $ \mu_{\mathbf{y}[m] \rightarrow \mathbf{x}[m]} $;
    \ENDFOR
\ENDFOR
\STATE {// \textit{Output a Posteriori Probability} //}
\FOR{$ m = 0 $ to $ N-1 $}
    \STATE using to \eqref{eq:pt}, compute $ \mathrm{Pr}^{ext}(\mathbf{x}[m]) $;
\ENDFOR
\STATE \textbf{Output:} $ \mathrm{Pr}^{ext}(\mathbf{x}_A,\mathbf{x}_B) $.

\end{algorithmic}
\end{algorithm}

\textit{Step 5 Output a Posteriori Probability:} Message passing between the nodes exploits both \textit{a priori} information and the likelihood function to progressively refine the reliability of the symbol estimates. After each message exchange, every variable node combines its \textit{a priori} information with all updated messages received from the connected factor nodes to compute the \textit{a posteriori} probabilities. For the $m$-th subcarrier, this update can be expressed as

\begin{equation}
\mathrm{Pr}^{ext}(\mathbf{x}[m]) \propto \mathrm{Pr}^{apr}(\mathbf{x}[m]) \prod_{n = m-D^m}^{m+D^m} \mu_{\mathbf{y}[n] \rightarrow \mathbf{x}[m]}.
\label{eq:pt}
\end{equation}

\textbf{Algorithm~\ref{alg:aca}} summarizes the complete procedure of the proposed adaptive channel-aware factor-graph detection (ACA-FGD) algorithm. A key advantage of ACA-FGD is its inherent parallelism, which enables multiple nodes to transmit messages to a common destination simultaneously, thereby reducing the overall processing latency.

\subsection{Parity-Check-Constraint-Based Soft-Information Refinement}

\begin{figure}[!t]
    \centering
    \includegraphics[width=1.0\linewidth]{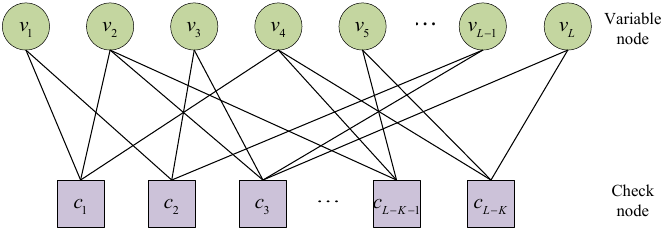}
    \caption{A Tanner graph of an LDPC code, illustrating the relationships between $L$ variable nodes and $L-K$ check nodes in a simplified manner.}
    \label{fig:tanner}
\end{figure}

Once the joint probability of the symbols $\mathrm{Pr}^{ext}(\mathbf{x}_A, \mathbf{x}_B)$ is obtained, demapping and deinterleaving are carried out. The \textit{a priori} probability $\mathrm{Pr}^{apr}(\mathbf{c}_A, \mathbf{c}_B)$ is then employed as the initial message for the variable nodes at the input of the decoder \cite{zhang2009channel}. In this work, we adopt the LDPC code of block length $L$ with $K$ information bits. The code is specified by its parity-check matrix $\mathbf{H} \in \{0, 1\}^{(L-K) \times L}$. The decoding is based on the generalized sum–product algorithm (G-SPA)~\cite{wubben2010generalized}, which leverages the linear nature of LDPC coding: the superposition of two separately encoded codewords from distinct nodes remains within the same code space, allowing joint decoding to process the combined signal. This iterative message-passing process is represented by a Tanner graph in Fig. \ref{fig:tanner}, where the structure of $\mathbf{H}$ defines the relationship between the variable nodes and the check nodes.

After each iteration of the message passing, the \textit{a posteriori} probability of the $i$-th variable node can be computed as $ \mathbf{p}_i[n]$, where $n \in\{0, 1, 2, 3 \}$ denotes the four possible joint bit states and $i \in \{0, 1, \cdots, L - 1\}$ indexes the variable nodes. Specifically, $ \mathbf{p}_i[0]=\mathrm{Pr}(\mathbf{c}_A[i]=0, \mathbf{c}_B[i]=0)$, $ \mathbf{P}_i[1]=\mathrm{Pr}(\mathbf{c}_A[i]=0, \mathbf{c}_B[i]=1)$, $ \mathbf{p}_i[2]=\mathrm{Pr}(\mathbf{c}_A[i]=1, \mathbf{c}_B[i]=0)$, $ \mathbf{p}_i[3]=\mathrm{Pr}(\mathbf{c}_A[i]=1, \mathbf{c}_B[i]=1)$. The hard decision for the $i$-th codeword $\mathbf{c}_R[i]$, is obtained by the PNC mapping rule.

\begin{equation}
\mathbf{c}_R[i] =
\begin{cases}
1 & \text{if } \underset{n}{\arg\max}~\mathbf{p}_i[n] = 1 \text{ or } 2, \\
0 & \text{others}.
\end{cases}
\end{equation}

The decoding process terminates when all parity-check equations are satisfied or the maximum number of iterations is reached. the parity-check syndrome is given by

\begin{equation}
\mathbf{s}^{(L-K) \times 1} = \mathbf{H}^{(L-K) \times L} \cdot \mathbf{c}_{R}^{~L \times 1}, \label{eq:syndrome}
\end{equation}
where $\mathbf{s} \in \{0, 1 \}$ and multiplication is over GF(2). When $\mathbf{s}[m]=1$ indicates that the $m$-th parity-check equation is violated, while $\mathbf{s}[m]=0$ denotes that it is satisfied. This allows the decoder to identify violated parity-check constraints and the associated variable nodes containing errors.

When all entries of $\mathbf{s}$ are zero, the decoding outputs $\mathbf{c}_R$. Otherwise, the reliability of the \textit{a posteriori} probability $ \mathbf{p}_i$ is taken as the \textit{a posteriori} probability $\mathrm{Pr}^{ext}(\mathbf{c}_A, \mathbf{c}_B)$, which is then fed back to the detector. The accuracy of the \textit{a priori} information plays a critical role in achieving reliable detection. Previous studies have validated that soft information can be refined during iterative processing \cite{ssettumba2023llr, porto2025iterative}. We adopt the general idea and extend it by incorporating parity-check constraints into the refinement.

\begin{algorithm}[!t]
\caption{Soft-Information Refinement Algorithm}
\label{alg:sr}
\begin{algorithmic}[1]

\STATE \textbf{Initiation:} $ \mathbf{z} = \mathbf{0}$;
\STATE \textbf{Input:} parity-check matrix $ \mathbf{H}$, parity-check syndrome $ \mathbf{s} $ and \textit{a posteriori} probability $ \mathrm{Pr}^{ext}(\mathbf{c}_A,\mathbf{c}_B) $;
\STATE {// \textit{Updating Belief Weights} //}
\FOR{$ k = 0 $ to $ L - K -1 $}
    \STATE find the corresponding set of variables: $\mathcal{N}(k)$;
    \FOR{$ i \in \mathcal{N}(k)$}
        \STATE perform \eqref{eq:z} to calculate $\mathbf{z}[i]$;
    \ENDFOR
\ENDFOR
\STATE {// \textit{Updating Refined Probability} //}
\FOR{$ i = 0 $ to $ L-1 $}
    \STATE compute \eqref{eq:w} to obtain $\mathbf{r}_{bf}[i]$;
    \STATE calculate $\mathrm{Pr}^{sr}(\mathbf{c}_A[i], \mathbf{c}_B[i])$ by \eqref{eq:sr};
\ENDFOR
\STATE \textbf{Output:} $\mathrm{Pr}^{sr}(\mathbf{c}_A, \mathbf{c}_B)$.

\end{algorithmic}
\end{algorithm}

In machine learning, belief scores are used to quantify the reliability of model predictions and to calibrate them into trustworthy probabilities~\cite{guo2017calibration, kuleshov2018accurate, minderer2021revisiting}. Motivated by this concept, we adopt a belief-weight mechanism to refine soft information, as outlined below.

\textit{Step 1 Initiation:} Let $\mathbf{z}$ denote the belief weight vector, which is initialized to the zero vector $\mathbf{0}$ at the beginning of the first detection and decoding iteration.

\textit{Step 2 Updating Belief Weights:} Denote the set of variable nodes connected to the $ k $-th check node by $\mathcal{N}(k)=\{ i: \mathbf{H}[k,i] = 1 \}$. The belief weights are updated according to \eqref{eq:z}, where $\alpha>0$ reinforces consistency when the $ k $-th syndrome bit satisfies $\mathbf{s}[k]=0$, and $\beta>0$ penalizes inconsistency when $\mathbf{s}[k]\neq 0$.

\begin{equation} \label{eq:z}
\mathbf{z}[i] =
\begin{cases}
\mathbf{z}[i] + \alpha, & \mathbf{s}[k] = 0, \\
\mathbf{z}[i] - \beta, & \mathbf{s}[k] \neq 0,
\end{cases}
\quad i \in \mathcal{N}(k).
\end{equation}

\textit{Step 3 Updating Refined Probability:} Updated belief weights are transformed into the probability domain through a \textit{sigmoid} function, ensuring that each belief probability remains within $[0,1]$:

\begin{equation} \label{eq:w}
\mathbf{r}_{bf}[i] = \frac{1}{1 + e^{-\mathbf{z}[i]}}.
\end{equation}

The resulting belief probability $\mathbf{r}_{bf}$ is then fused with the \textit{a posteriori} probability $\mathrm{Pr}^{ext}(\mathbf{c}_A,\mathbf{c}_B)$ to generate a reinforced probability $\mathrm{Pr}^{sr}(\mathbf{c}_A,\mathbf{c}_B)$. To preserve probabilistic consistency, a uniform compensation term $\mathbf{1}/4$ and a normalization factor $\xi$ are incorporated, which slightly moderates the update strength. For the $i$-th variable node, the refinement step is expressed as
\begin{equation} \label{eq:sr}
\mathrm{Pr}^{sr}(\mathbf{c}_A[i],\mathbf{c}_B[i])
= \xi \left(
\mathbf{r}_{bf}[i] \, \mathrm{Pr}^{ext}(\mathbf{c}_A[i],\mathbf{c}_B[i])
+ \mathbf{1}/{4}
\right).
\end{equation}

In summary, the proposed algorithm exploits parity-check constraints to reinforce or attenuate \textit{a posteriori} information, thereby enhancing the accuracy of \textit{a priori} knowledge provided to the detection. The complete refinement process is presented in \textbf{Algorithm~\ref{alg:sr}}.

\subsection{Superimposed-Signal-Based Linear Solution}

Although the proposed ACA-FGD achieves optimal performance, the intrinsic complexity of the factor graph prevents it from being lightweight. In scenarios where processing latency is more important than detection accuracy, it may lose its advantage. Therefore, we derive a closed-form and lower-complexity linear solution that is suited for cases with modest performance requirements. In the multi-hop relay setting, the received signal is treated as a superimposed statistic $\mathbf{x}$. It is just introduced to characterize the superposition structure of the received waveform. Although the linear estimation is derived from the superimposed model, the soft-output computation is still performed over the joint hypothesis $(\mathbf{x}_A,\mathbf{x}_B)$.

\begin{equation}
\mathbf{x} = \mathbf{x}_A + \mathbf{x}_B.
\end{equation}

Considering that our method is based on an iterative refinement of the prior mean and variance of $ \mathbf{x}_u $, we denote them by $ \bar{\mathbf{x}}_u $ and $ \mathbf{V}_u = \mathrm{diag}(\mathbf{v}_u) $, respectively. Due to this, the derivation of linear detection can be given as follows.

\textit{Step 1 Initiation:} We define $\mathrm{Pr}^{apr}(\mathbf{x}_u)$ as the \textit{a priori} probability. During the first linear detection of the received signal, it is initialized to a uniform distribution vector $\mathbf{1}/{2^{Q}}$.

\textit{Step 2 Depth Selection:} Full-matrix operations dominate the computational cost of linear detection. Similar to the adaptive strategy, we simplify $\mathbf{H}_u$ according to \eqref{eq:eta_tD} and \eqref{eq:Dm}.

\textit{Step 3 Updating a Priori Information:} When linear detection obtains \textit{a priori} information from the previous iteration, it updates $ \bar{\mathbf{x}}_{u} $ and $ \mathbf{v}_{u} $ by

\begin{align}
\mathrm{Pr}^{apr}(\mathbf{x}_u) &=  \sum_{ \sim (\mathbf{x}_u)} \mathrm{Pr}^{apr}(\mathbf{x}_A,\mathbf{x}_B), \label{eq:p} \\
\bar{\mathbf{x}}_u &= \sum_{i=1}^{2^Q} \mathcal{L}_u[i] \mathrm{Pr}^{apr}(\mathbf{x}_u = \mathcal{L}_u[i]), \label{eq:x} \\
\mathbf{v}_u &= \sum_{i=1}^{2^Q} \left\lVert \mathcal{L}_u[i] - \bar{\mathbf{x}}_u \right\lVert^2 \mathrm{Pr}^{apr}(\mathbf{x}_u = \mathcal{L}_u[i]). \label{eq:v}
\end{align}

\textit{Step 4 Derivation of the Superimposed Model:} Based on the MMSE principle\cite{tuchler2002turbo}, the estimation $ \hat{\mathbf{x}} $ can be given by

\begin{equation}
\begin{aligned}
\hat{\mathbf{x}} &= \mathbb{E}[\mathbf{x}] + \mathbf{R}_{\mathbf{xy}} \mathbf{R}_{\mathbf{yy}}^{-1} ( \mathbf{y} - \mathbb{E}[\mathbf{y}] )\\
&= \bar{\mathbf{x}}_{A} + \bar{\mathbf{x}}_{B} + (\mathbf{V}_A \mathbf{H}_A^H + \mathbf{V}_B \mathbf{H}_B^H) (\mathbf{H}_A \mathbf{V}_A \mathbf{H}_A^H \\
&+ \mathbf{H}_B \mathbf{V}_B \mathbf{H}_B^H + \sigma^2 \mathbf{I}_N)^{-1} (\mathbf{y} - \mathbf{H}_{A} \bar{\mathbf{x}}_A -  \mathbf{H}_{B} \bar{\mathbf{x}}_B).
\end{aligned}
\label{eq:lmmse}
\end{equation}

\begin{figure*}[!t]
    \centering
    \subfloat[UWA CIRs when $ \sigma_{u} = 0.1 $ m/s]{\includegraphics[width=0.49\linewidth]{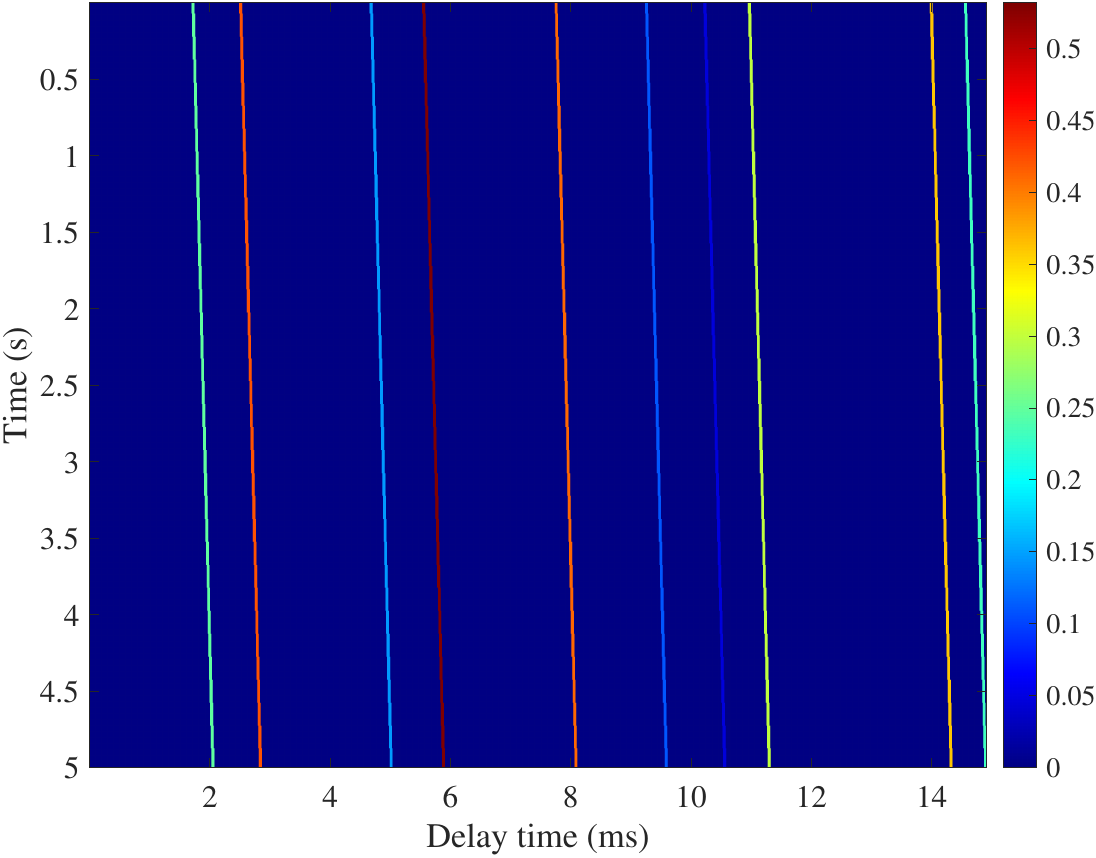}\label{fig:sigma01}}
    \hfill
    \subfloat[UWA CIRs when $ \sigma_{u} = 1.5 $ m/s]{\includegraphics[width=0.49\linewidth]{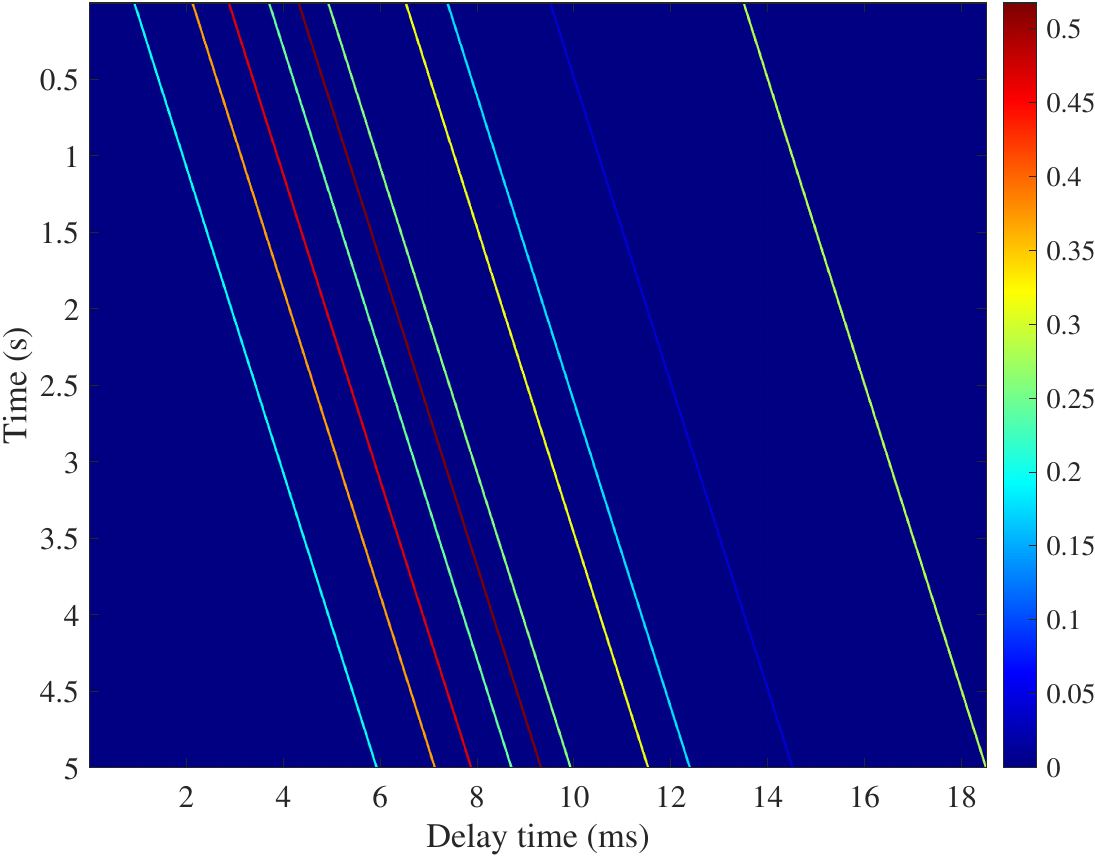}\label{fig:sigma15}}
    \caption{UWA CIRs with fixed deviated speed $ \sigma_{u}$ which correspond to fixed Doppler factors of $ a_{p,u} = 6.77 \times 10^{-5} $ and $ 1.00 \times 10^{-3} $.}
    \label{fig:uwa_doppler}
\end{figure*}

\begin{algorithm}[!t]
\caption{Proposed SM-LMMSE Algorithm}
\label{alg:mmse}
\begin{algorithmic}[1]

\STATE \textbf{Initiation:} $ \mathrm{Pr}^{apr} (\mathbf{x}_{u}) = \mathbf{1}/2^Q $;
\STATE \textbf{Input:} received frequency-domain vector $ \mathbf{y} $, frequency-domain channel matrices $ \mathbf{H}_{A} $ and $ \mathbf{H}_B $, noise power $ \sigma^2 $, \textit{a priori} probability $ \mathrm{Pr}^{apr}(\mathbf{x}_A,\mathbf{x}_B) $;
\STATE {// \textit{Depth Selection} //}
\IF{first detection in the iterative receiver}
    \FORALL{ $ m = 0 $ to $ N-1 $}
        \STATE combine \eqref{eq:eta_tD} to determine $D_{u}^{m,*}$;
        \STATE perform \eqref{eq:Dm} to choose $ D^m $;
    \ENDFOR
\ENDIF
\STATE {// \textit{Updating a Priori Information} //}
\IF {obtain $ \mathrm{Pr}^{apr}(\mathbf{x}_A,\mathbf{x}_B) $ from the previous iteration}
    \STATE update $ \bar{\mathbf{x}}_u $ and $ \mathbf{v}_{u} $ by \eqref{eq:x} and \eqref{eq:v};
\ENDIF
\STATE {// \textit{Derivation of the Superimposed Model} //}
\STATE compute $ \hat{\mathbf{x}} $ by \eqref{eq:MMSExx};
\STATE {// \textit{Conditional Probability with Mean and Variance} //}
\FOR{$ i = 0 $ to $ 4^{Q}-1 $}
    \STATE update  $ \mu_{\mathbf{x}}[i] $ and $ \sigma_{\mathbf{x}}^2[i] $ from \eqref{eq:MMSEu} and \eqref{eq:MMSEsigma};
    \STATE perform \eqref{eq:MMSEP} to obtain $  \mathrm{Pr}(\hat{\mathbf{x}} \mid \mathbf{x} = \mathcal{L}[i]) $;
\ENDFOR
\STATE \textbf{Output:} $ \mathrm{Pr}^{ext}(\mathbf{x}_A, \mathbf{x}_B) $.

\end{algorithmic}
\end{algorithm}

However, $ \hat{\mathbf{x}} $ is dependent on \textit{a priori} information from $ \mathbf{x}_A $ and $ \mathbf{x}_B $. To decouple from previous knowledge, the prior mean and variance of $ \mathbf{x}_u $ are compensated for $ \mathbf{0} $ and $ \mathbf{1} $. Under this assumption, \eqref{eq:lmmse} can be reformulated as

\begin{equation}
\begin{aligned}
\hat{\mathbf{x}} &= (\mathbf{S}_{A}^H + \mathbf{S}_{B}^H) ( \sum_{u \in \{ A,B \}} \mathbf{H}_{u} \mathbf{V}_u \mathbf{H}_{u}^H + \mathbf{S}_{u} (\mathbf{I}_N-\mathbf{V}_u) \mathbf{S}_{u}^H \\
&+ \sigma ^2 \mathbf{I}_N )^{-1} ( \mathbf{y} - \sum_{u \in \{ A,B \}} \left( \mathbf{H}_{u} \bar{\mathbf{x}}_u - \mathbf{S}_{u} \bar{\mathbf{x}}_u \right) ),
\label{eq:MMSExx}
\end{aligned}
\end{equation}
where $ \mathbf{S}_{u} = \mathrm{diag}(\mathbf{H}_{u}) $ shows a simplified channel to emphasize that each subcarrier makes its dominant contribution to the corresponding position in the received subcarrier. To simplify \eqref{eq:MMSExx}, we define the coefficient matrix of the linear method as $\mathbf{C} = \mathbf{R}_{\mathbf{xy}} \mathbf{R}_{\mathbf{yy}}^{-1}$.

\textit{Step 5 Conditional Probability with Mean and Variance:} To derive the soft input soft output solution, $ \hat{\mathbf{x}} $ is modeled as a Gaussian random variable conditioned on the superimposed symbol $ \mathcal{L} = \mathcal{L}_{A} + \mathcal{L}_{B} $. Consequently, the \textit{a posteriori} output $ \mathrm{Pr}^{ext}(\mathbf{x}_A, \mathbf{x}_B) $ can be characterized by conditional mean and variance. Let $ l_u $ denote a possible constellation value of $ \mathcal{L}_u $ and then the conditional mean $ \mu_{\mathbf{x}} $ can be expressed as

\begin{equation}
\begin{aligned}
\mu_{\mathbf{x}}[i] &= \mathbb{E}[\hat{\mathbf{x}} \mid \mathbf{x} = \mathcal{L}[i]] \\
&= \mathbf{C} ( \mathbb{E}[\mathbf{y} \mid \mathbf{x} = \mathcal{L}[i]] - \mathbb{E}(\mathbf{y}) ) \\
&=  \mathbf{C} \left( l_A \cdot \mathbf{S}_A + l_B \cdot \mathbf{S}_B \right),
\label{eq:MMSEu}
\end{aligned}
\end{equation}
where $ i \in \{ 0, 1, \cdots, 4^{Q}-1 \} $. Similarly, the conditional variance $ \sigma_{\mathbf{x}}^2 $ is given by

\begin{equation}
\begin{aligned}
\sigma_{\mathbf{x}}^2[i] &= \mathbb{E}[\left\lVert \hat{\mathbf{x}} - \mu_{\mathbf{x}}[i] \right\lVert ^2 \mid \mathbf{x} = \mathcal{L}[i] ] \\
&= \mathbf{C}  \mathbf{R}_{[\mathbf{yy} \mid \mathbf{x} = \mathcal{L}[i]} \mathbf{C}^H \\ 
&= \mathbf{C} ( \sigma^2 \mathbf{I}_N + \sum_{u \in {A,B}} \mathbf{H}_{u} \mathbf{V}_u \mathbf{H}_{u}^H - \mathbf{S}_{u} \mathbf{V}_u \mathbf{S}_{u}^H) \mathbf{C}^H. \label{eq:MMSEsigma} \\
\end{aligned}
\end{equation}

Due to the above derivations and assumptions, the conditional probability output can be shown as

\begin{figure*}[!t]
    \centering
    \subfloat[BER comparison in the $ \sigma_{u} = 0.1 $ m/s UWA channel (single relay)]{\includegraphics[width=0.49\linewidth]{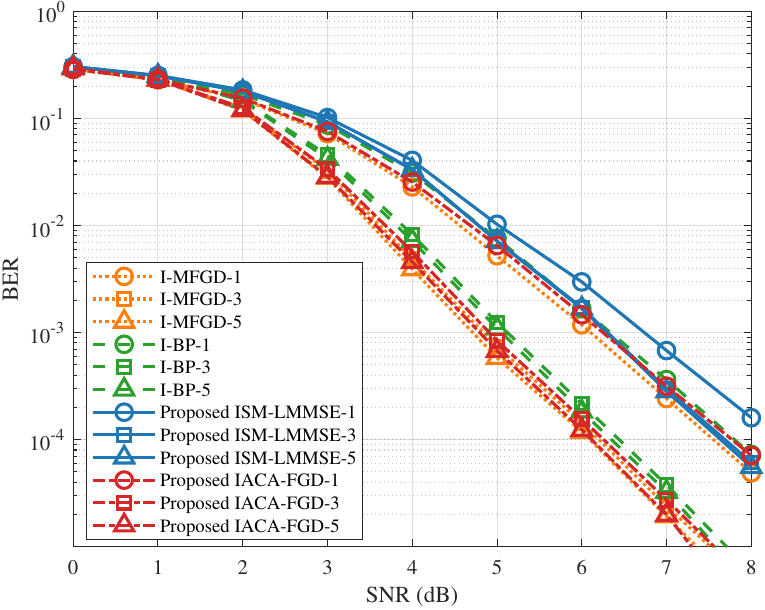}\label{fig:CF1}}
    \hfill
    \subfloat[BER comparison in the $ \sigma_{u} = 1.5 $ m/s UWA channel (single relay)]{\includegraphics[width=0.49\linewidth]{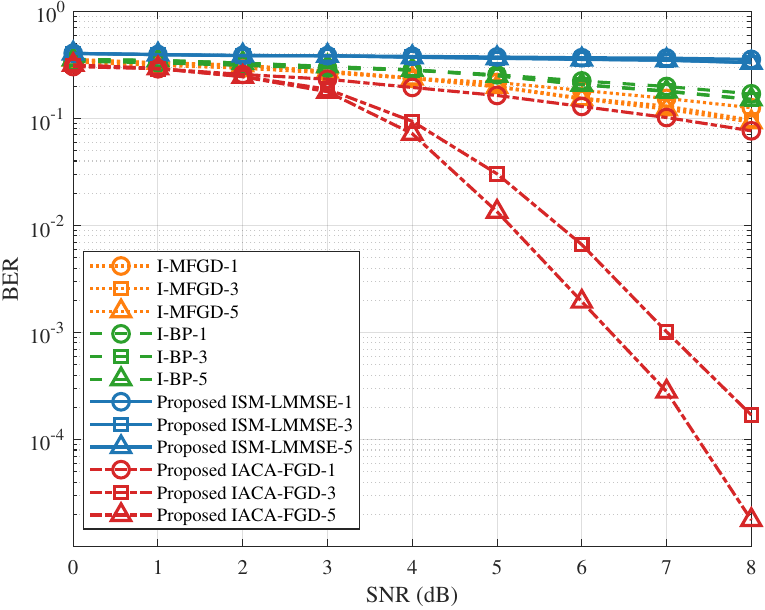}\label{fig:CF2}}
    \vspace{-1mm}  
    \subfloat[BER comparison in the $ \sigma_{u} = 0.1 $ m/s UWA channel (multiple relays)]{\includegraphics[width=0.49\linewidth]{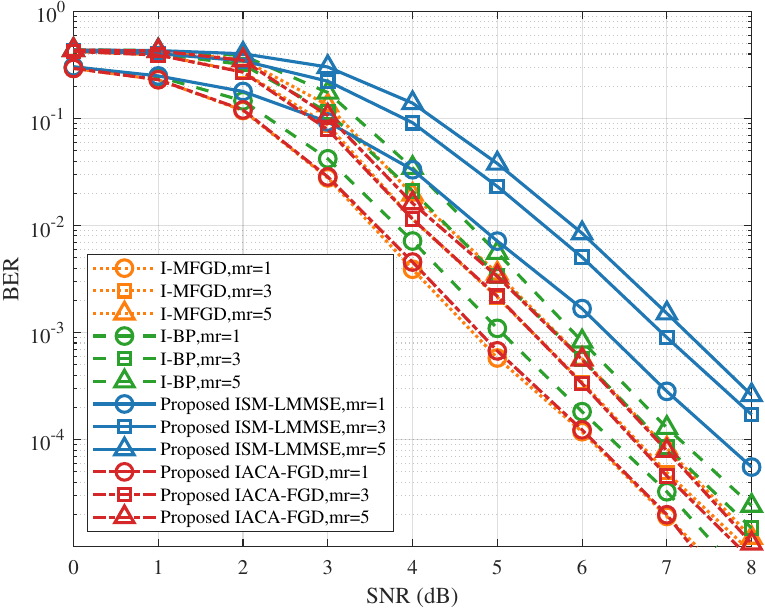}\label{fig:CF3}}
    \hfill
    \subfloat[BER comparison in the $ \sigma_{u} = 1.5 $ m/s UWA channel (multiple relays)]{\includegraphics[width=0.49\linewidth]{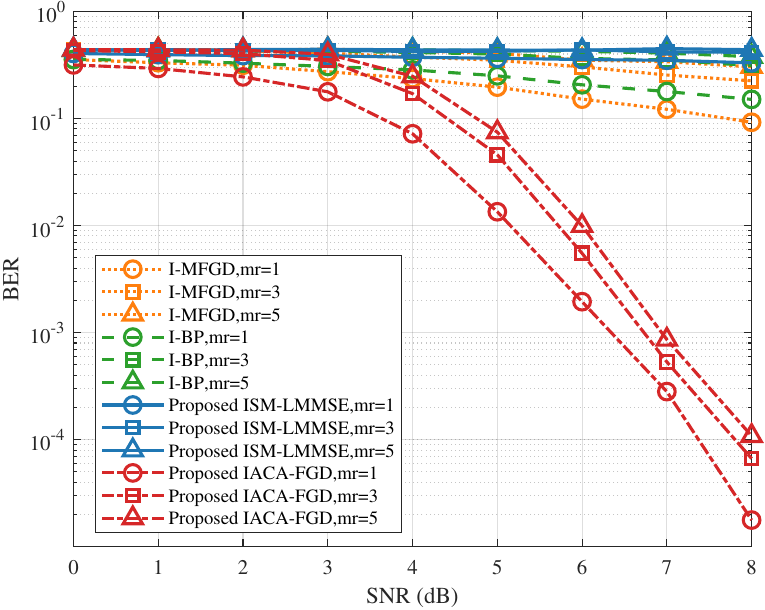}\label{fig:CF4}}
    \caption{Performance comparison of different iterative receiver schemes with a number of outer iterations and multi-hop transmission in UWA channels.}
    \label{fig:CF}
\end{figure*}

\begin{equation}
\mathrm{Pr}(\hat{\mathbf{x}} \mid \mathbf{x} = \mathcal{L}[i]) \propto \exp\left( - \frac{ \left\lVert \mu_{\mathbf{x}}[i] - \hat{\mathbf{x}} \right\lVert^2 }{\sigma_{\mathbf{x}}^2[i]} \right). \label{eq:MMSEP}
\end{equation}

\begin{table}[!t]
    \renewcommand{\arraystretch}{1.2}
    \centering
    \caption{Computation Complexity of Proposed Detection Algorithms}
    \label{tab:cc}
    \begin{tabular}{lccc}
        \toprule
        \textbf{Algorithm} & \textbf{Computation Complexity} \\
        \midrule
        ACA-FGD & $\mathcal{O}(N(D^m)^2 4^{Q D^m})$ \\
        SM-LMMSE & $\mathcal{O}(N(D^m + 4^Q))$ \\
        \bottomrule
    \end{tabular}
\end{table}

Since its derivation is rooted in the superposed model, we refer to SM-LMMSE. The complete procedure is summarized in \textbf{Algorithm~\ref{alg:mmse}}. We evaluate the computational complexity of the proposed detection algorithms in Table~\ref{tab:cc}. As $D^{m}$ and $Q$ increase, the computational complexity of SM-LMMSE grows linearly, whereas that of ACA-FGD increases exponentially. Under high-order modulation and complex channel conditions, linear detection can offer a faster execution speed. As depth selection has virtually no impact on linear detection performance, it is not considered in subsequent simulations and experiments. The corresponding analysis follows ACA-FGD.

\section{Simulation Results}\label{sec:simulation}

\begin{figure*}[!t]
    \centering
    \subfloat[BER of different \emph{ICI depths} in the $ \sigma_{u} = 0.1 $ m/s UWA channel]{\includegraphics[width=0.49\linewidth]{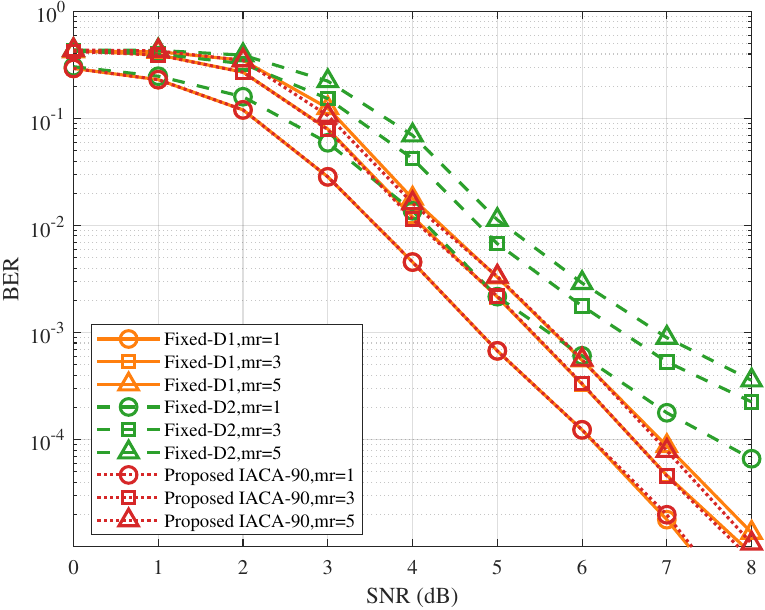}\label{fig:ICI01}}
    \hfill
    \subfloat[BER of different \emph{ICI depths} in the $ \sigma_{u} = 0.5 $ m/s UWA channel]{\includegraphics[width=0.49\linewidth]{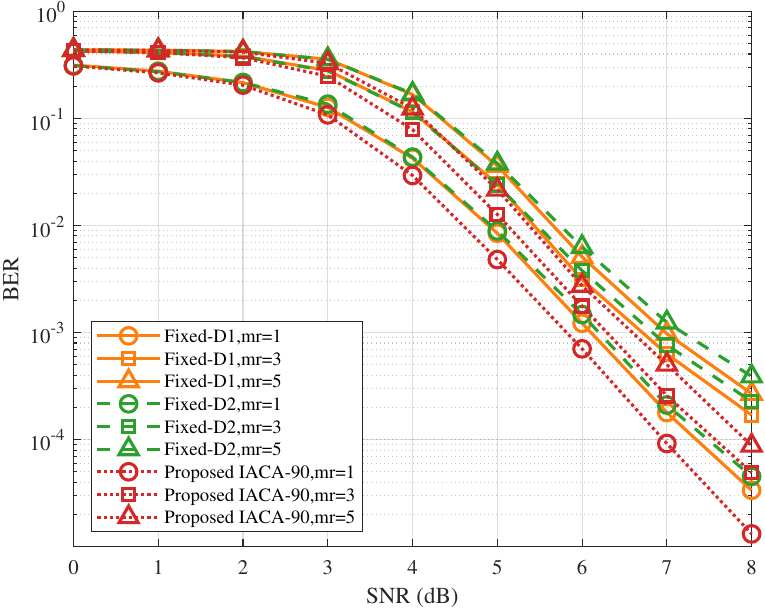}\label{fig:ICI05}}
    \vspace{-1mm}  
    \subfloat[BER of different \emph{ICI depths} in the $ \sigma_{u} = 1.0 $ m/s UWA channel]{\includegraphics[width=0.49\linewidth]{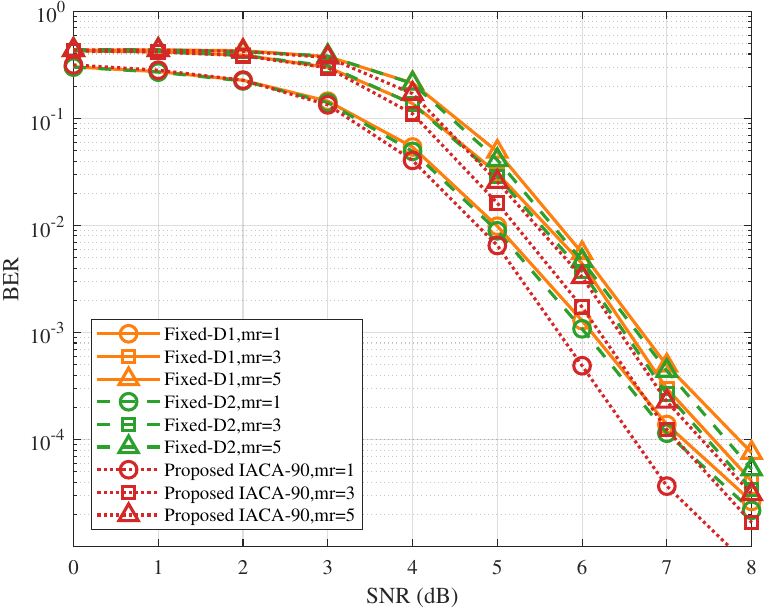}\label{fig:ICI10}}
    \hfill
    \subfloat[BER of different \emph{ICI depths} in the $ \sigma_{u} = 1.5 $ m/s UWA channel]{\includegraphics[width=0.49\linewidth]{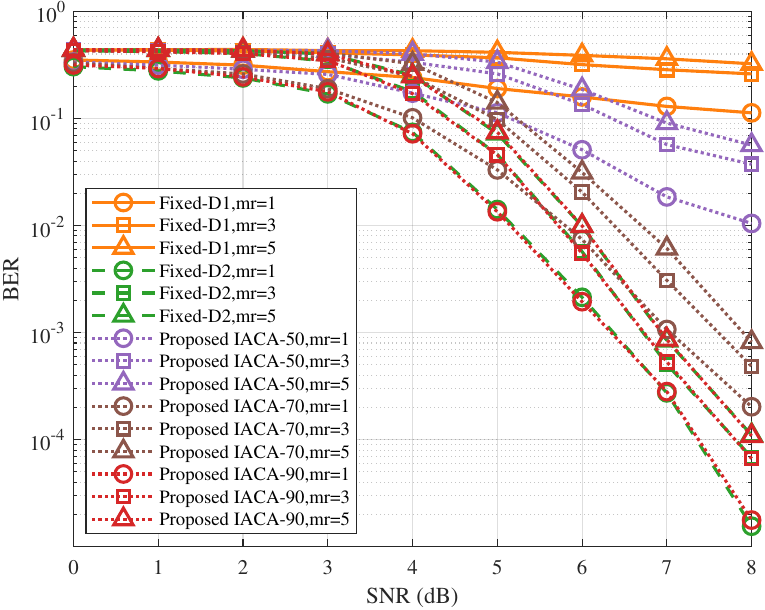}\label{fig:ICI15}}
    \caption{Performance of different \emph{ICI depths} in the  UWA channels. To distinguish \emph{ICI depths} strategies, we denote the proposed method as IACA, while the traditional fixed \emph{ICI depth} approach is defined as Fixed-D. The thresholds for IACA-FGD are marked as 60\%, 70\%, 80\% and 90\%.}
    \label{fig:ICI}
\end{figure*}

In this section, we evaluate the performance of the proposed algorithms through simulations. The multi-hop UWA PNC system employs OFDM modulation, with carrier frequency $ f_c = 22.4 $ kHz, bandwidth $ B \approx 5 $ kHz, sampling rate $ f_s = 50 $ kHz, and symbol duration $ T = 102.4 $ ms including cyclic prefix $ T_{cp} = 20.5 $ ms. We design 336 subcarrier positions for data transmission with a rate-$ 1/3 $ irregular QC-LDPC code over $ \mathrm{GF}(2) $, followed by BPSK modulation. We name the iterative receiver schemes based on the proposed algorithms as IACA-FGD and ISM-LMMSE. To highlight the effectiveness of designed schemes, we select representative schemes, where the detection algorithms are from \cite{situ2018ofdm} and \cite{zhou2014ofdm}, mentioned as I-BP and I-MFGD, respectively. Under normal conditions, the outer iteration step includes 1 signal detection and 3 channel decoding. In addition, we consider up to $mr=5$ relay nodes in the multi-hop network.

The relevant parameters of the UWA channel are configured according to \cite{wang2013iterative}. We assume that the channel consists of 10 independent paths. The arrival times of these paths follow an exponential distribution with an average delay of $ 1 $ ms. The amplitude gain of each path is Rayleigh distributed and the average power of each path decays exponentially with delay, resulting in a 20 dB difference between the start and end of the CP. The Doppler factor governs the temporal variation of the MSML model, where the factor for each path is drawn from a 0 mean uniform distribution parameterized by the velocity deviation $ \sigma_{u} $.

It is assumed that the UWA channels experienced by the messages transmitted from the uplink node $A$ and the downlink node $B$ to the relay node $R$ share identical parameters. Fig.~\ref{fig:uwa_doppler} illustrates the UWA CIRs at a fixed $\sigma_{u}$, revealing the coupling between time, delay and amplitude. When $\sigma_{u}$ is small, the UWA channel exhibits a mild delay spread. As $\sigma_{u}$ increases to 1.5 m/s, the nature of the time-varying channel becomes more remarkable, with a delay span of approximately 3-4 ms.

\subsection{Performance Comparison with Benchmark Schemes}

\begin{figure*}[!t]
    \centering
    \subfloat[BER under ISM-LMMSE scheme]{\includegraphics[width=0.49\linewidth]{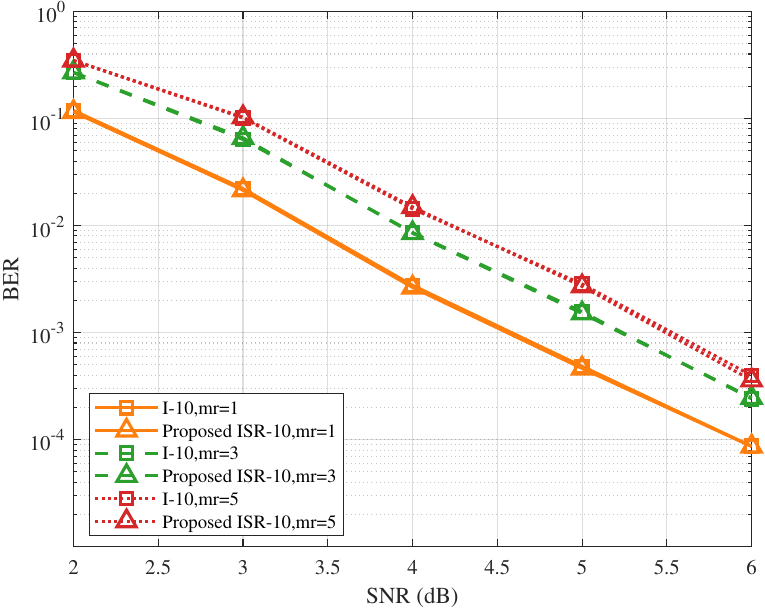}\label{fig:BF-LMMSE}}
    \hfill
    \subfloat[BER under I-MFGD scheme]{\includegraphics[width=0.49\linewidth]{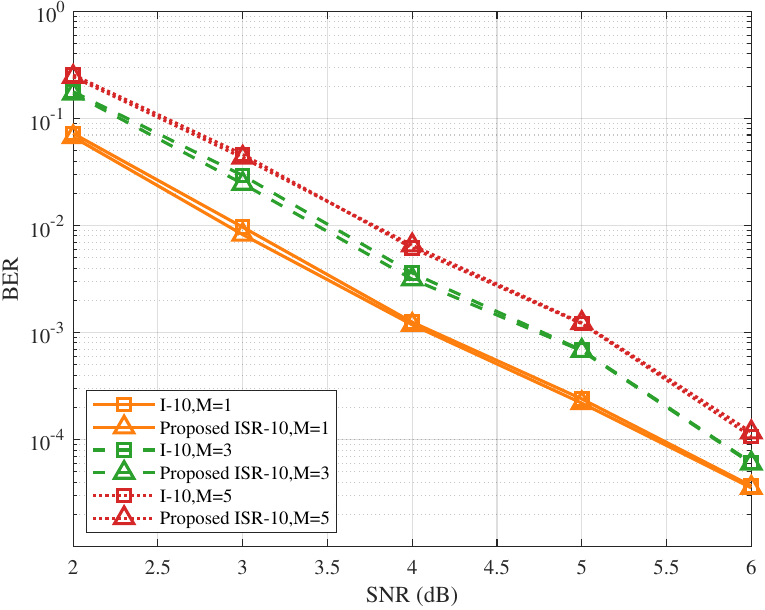}\label{fig:BF-MFGD}}
    \vspace{-1mm}  
    \subfloat[BER under I-BP scheme]{\includegraphics[width=0.49\linewidth]{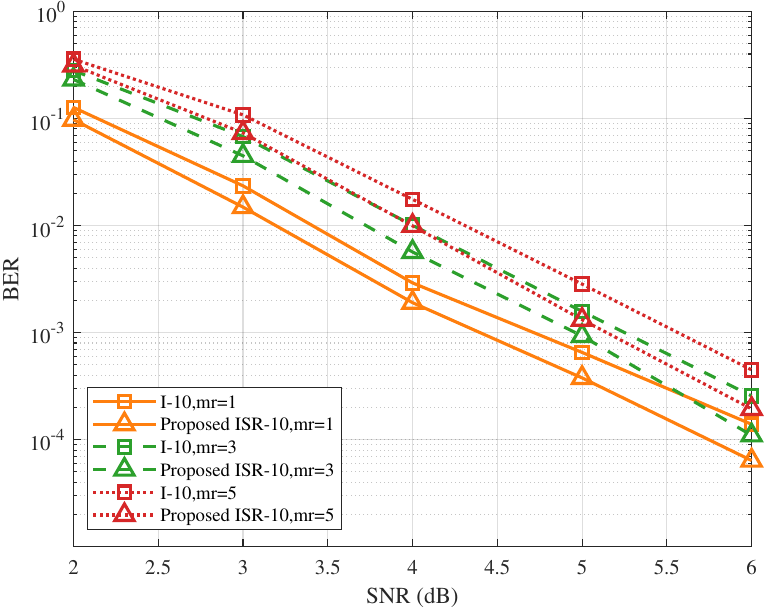}\label{fig:BF-BP}}
    \hfill
    \subfloat[BER under IACA-FGD scheme]{\includegraphics[width=0.49\linewidth]{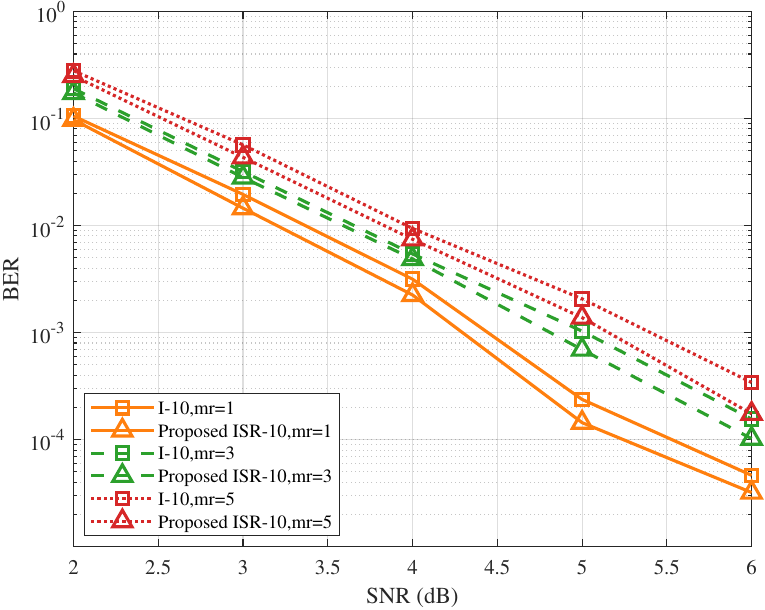}\label{fig:BF-FGD}}
    \caption{Performance comparison of soft-information refinement in the $ \sigma_{u} = 0.1 $ m/s UWA channel.}
    \label{fig:BER-BF}
\end{figure*}

Fig.~\ref{fig:CF} compares different signal processing schemes within the classical iterative framework. For a practical comparison, \emph{ICI depth} in both I-MFGD and I-BP is fixed at $D^{m}=1$, and the adaptive threshold $\eta_{u}^{m}$ in IACA-FGD is set to $90\%$.

Figs.~\ref{fig:CF1} and \ref{fig:CF2} present the performance of different schemes in a single relay transmission setting with the maximum number of outer iterations set to $1$, $3$, and $5$. It is evident that the iterative receiver framework yields performance gains in UWA channels, achieving convergence after approximately 5 outer iterations, which underscores the effectiveness in exploiting \textit{a priori} information fed back from the decoder. In the $\sigma_{u} = 0.1$~m/s channel, the MP-based schemes exhibit comparable performance, approaching a BER of $10^{-4}$ with a 2~dB gain from the outer iteration, while the iterative gain achieved by ISM-LMMSE is limited to less than 1~dB. Due to its inability to exploit inter-node dependencies, ISM-LMMSE achieves limited performance. 

When $\sigma_{u} = 1.5$~m/s, most schemes fail to perform effective detection. Only the proposed IACA-FGD achieves reliable performance, reaching a BER close to $10^{-5}$ at an SNR of 8~dB. Considering that non-iterative cases of all schemes tend to saturate around a BER of $10^{-1}$, iterative processing becomes indispensable in highly dynamic channels.

Figs.~\ref{fig:CF3} and~\ref{fig:CF4} consider multi-hop forwarding with up to five relays. For each relay, the outer iteration is set to $5$, and the performance under $mr\in\{1,3,5\}$ is reported. When $\sigma_u=0.1$~m/s, MP-based schemes exhibit stable performance, with BER degradation less than $1$~dB after 5 hop forwarding. The BER of ISM-LMMSE degrades more noticeably, but it approaches BER $=10^{-4}$ at $\mathrm{SNR}=8$~dB. In the $\sigma_u=1.5$~m/s channel, IACA-FGD maintains a meaningful performance, while the other schemes fail completely.

\subsection{Impact of Adaptive Channel-Aware Detection}

\begin{figure*}[!t] 
    \centering
    \subfloat[BER of different CERs in the $ \sigma_{u} = 0.1 $ m/s UWA channel]{\includegraphics[width=0.49\linewidth]{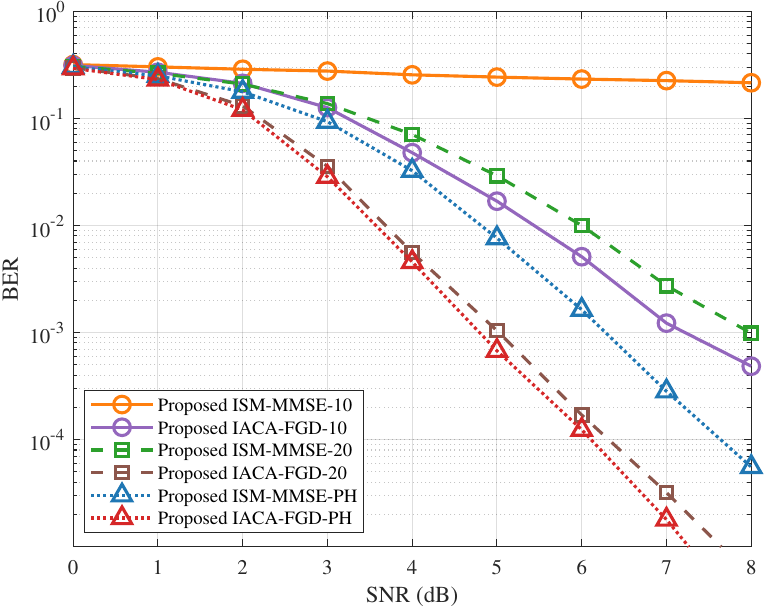}\label{fig:CER-0.1}}
    \hfill
    \subfloat[BER of different CERs in the $ \sigma_{u} = 1.5 $ m/s UWA channel]{\includegraphics[width=0.49\linewidth]{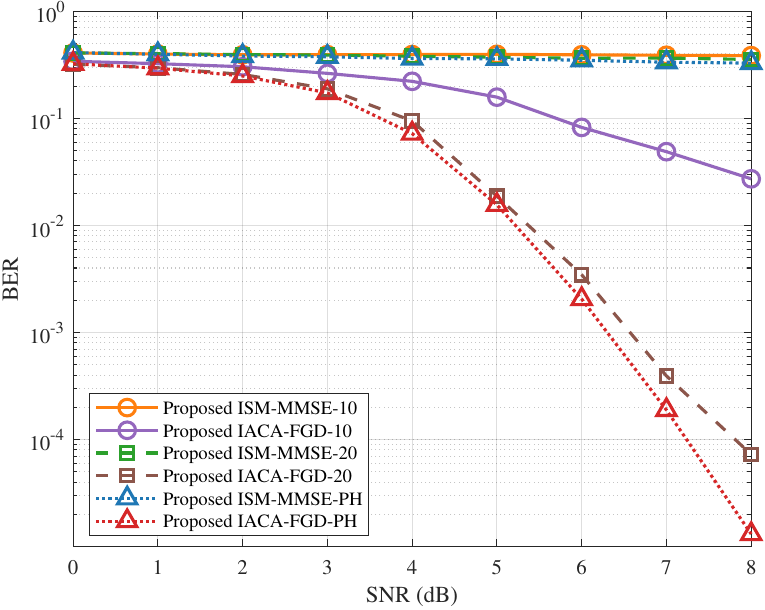}\label{fig:CER-1.5}}
    \caption{Performance comparison of different CERs in the UWA channels. System simulation is conducted under CER $ = 10 $ and $ 20 $ dB, with comparisons drawn against the perfect channel (PH) baseline.}
    \label{fig:BER-CER}
\end{figure*}

As shown in Table~\ref{tab:energy_contribution}, an average energy analysis is conducted over 30,000 samples across different channels. We assume that all subcarrier positions have the same \emph{ICI depth} in the calculations, providing a clear representation of the channel energy distribution and the sidelobe contribution. When the velocity deviation is small, most of the channel energy remains concentrated along the main diagonal, and a depth of $D^m=1$ is sufficient to characterize the UWA channels. As $\sigma_{u}$ increases, the energy becomes more dispersed and The necessity of extending the detection range must be considered. When $\sigma_{u}=1.5$~m/s, the cumulative ratio exceeds 90\% at $D^m=2$.

\begin{table}[htbp]
    \renewcommand{\arraystretch}{1.2}
    \centering
    \caption{Channel Energy Contribution under Different \emph{ICI depths}}
    \label{tab:energy_contribution}
    \begin{tabular}{lccc}
        \toprule
        \textbf{Doppler Rate} & \textbf{$ D^m = 0 $} & \textbf{$ D^m = 1 $} & \textbf{$ D^m = 2 $} \\
        \midrule
        $ \sigma_{u} = 0.1 $ m/s & 98.40\% & 99.42\% & 99.67\% \\
        $ \sigma_{u} = 0.5 $ m/s & 69.68\% & 95.89\% & 97.90\% \\
        $ \sigma_{u} = 1.0 $ m/s & 37.65\% & 93.85\% & 96.84\% \\
        $ \sigma_{u} = 1.5 $ m/s & 26.34\% & 77.53\% & 95.18\% \\
        \bottomrule
    \end{tabular}
\end{table}

Fig.~\ref{fig:ICI} illustrates the receiver performance under different \emph{ICI depths}, with a focus on the results with up to $mr=5$ relays and 5 outer iterations. When $\sigma_{u}=0.1$~m/s, Fixed-D1 converges almost 2~dB faster than Fixed-D2 at a BER of $10^{-4}$. Their performance gap varies as the sidelobe energy increases. When $\sigma_u=0.5$~m/s, Fixed-D1 slightly outperforms Fixed-D2, whereas at $\sigma_u=1.0$~m/s, Fixed-D2 becomes marginally superior to Fixed-D1.

Given the large discrepancy observed at $\sigma_{u}=1.5$~m/s, where the channel energy becomes dispersed, we further investigate the optimal adaptive threshold. It is obvious that IACA-90 achieves a performance comparable to Fixed-D2 with a BER close to $10^{-4}$ at 8~dB SNR after 5 relays transmission. Considering the situations of all threshold settings, we select $\eta_{u}^{m}=90\%$ as the optimal configuration. In the $\sigma_{u}=0.5$ and $1.0$~m/s channels, IACA-90 outperforms both fixed schemes by approximately 1~dB, indicating that the adaptive strategy captures the channel structure and thus enables more reliable detection.

\subsection{Impact of Soft-Information Refinement in the Iterative Receiver}

To highlight the performance gain from soft information refinement, we choose the $\sigma_{u} = 0.1$ channel for comparison, as it provides a reasonable baseline performance even without the above approach. Since the number of decoding iterations affects the performance of refinement gain, we focus on the results for 10 decoding iterations with the maximum number of outer iterations set to $5$ with $mr=5$ relays. Given that parity-check errors are critical to the decoding outcome, we set the parameters to $\alpha = 1$ and $\beta = 5$. For clarity, the classical iterative framework is denoted as I, while the framework with soft-information refinement is denoted as ISR.

Figs.~\ref{fig:BER-BF}\subref{fig:BF-LMMSE} and \ref{fig:BER-BF}\subref{fig:BF-MFGD} demonstrate that soft-information refinement results in little performance improvement across various decoding iterations. This is due to two factors: the simple computation process of ISM-LMMSE leads to limited performance gains from the \textit{a priori} knowledge, while the Markov model shows low sensitivity to \textit{a priori} information.

In contrast, Figs.~\ref{fig:BER-BF}\subref{fig:BF-BP} and \ref{fig:BER-BF}\subref{fig:BF-FGD} present the advantages of soft-information refinement. It can be obvious that the refinement gains in I-BP and IACA-FGD achieve about 0.5~dB and 0.2~dB after 5 relays transmission. The reason why the results in I-BP are different from I-MFGD is that although it employs a Markov structure, its localized computation limits performance, a drawback that is mitigated through soft-information refinement. Factor graph detection outperforms the Markov model in leveraging \textit{a priori} information by representing the relationships between the variable nodes and the factor nodes.

\subsection{Robustness to Channel Estimation Errors}

\begin{figure}[!t]
    \centering
    \subfloat[Furong Lake environment]{\includegraphics[width=0.9\linewidth]{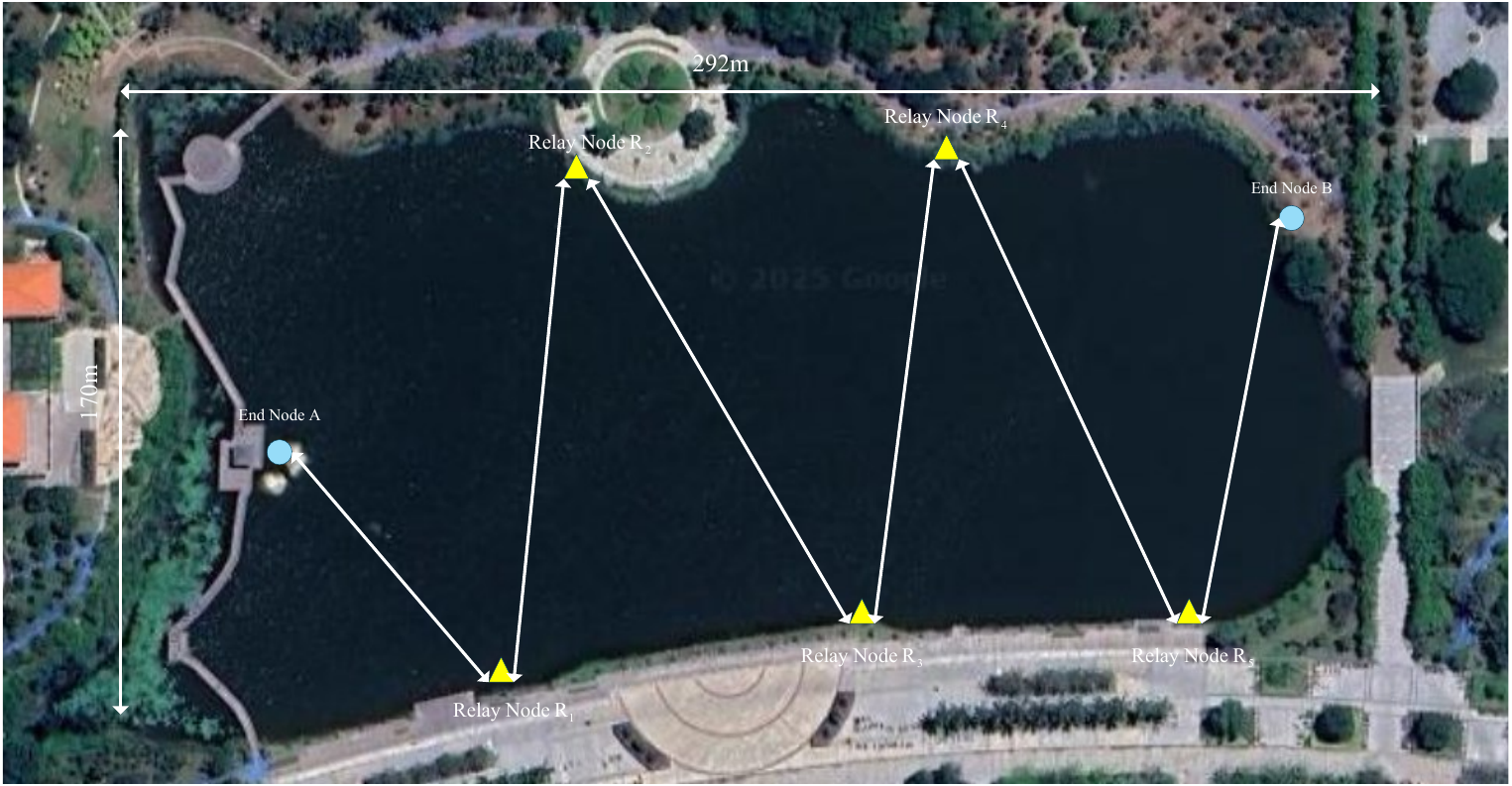}\label{fig:scenario_Lake}}
    \hfill
    \subfloat[Taiwan Strait environment]{\includegraphics[width=0.9\linewidth]{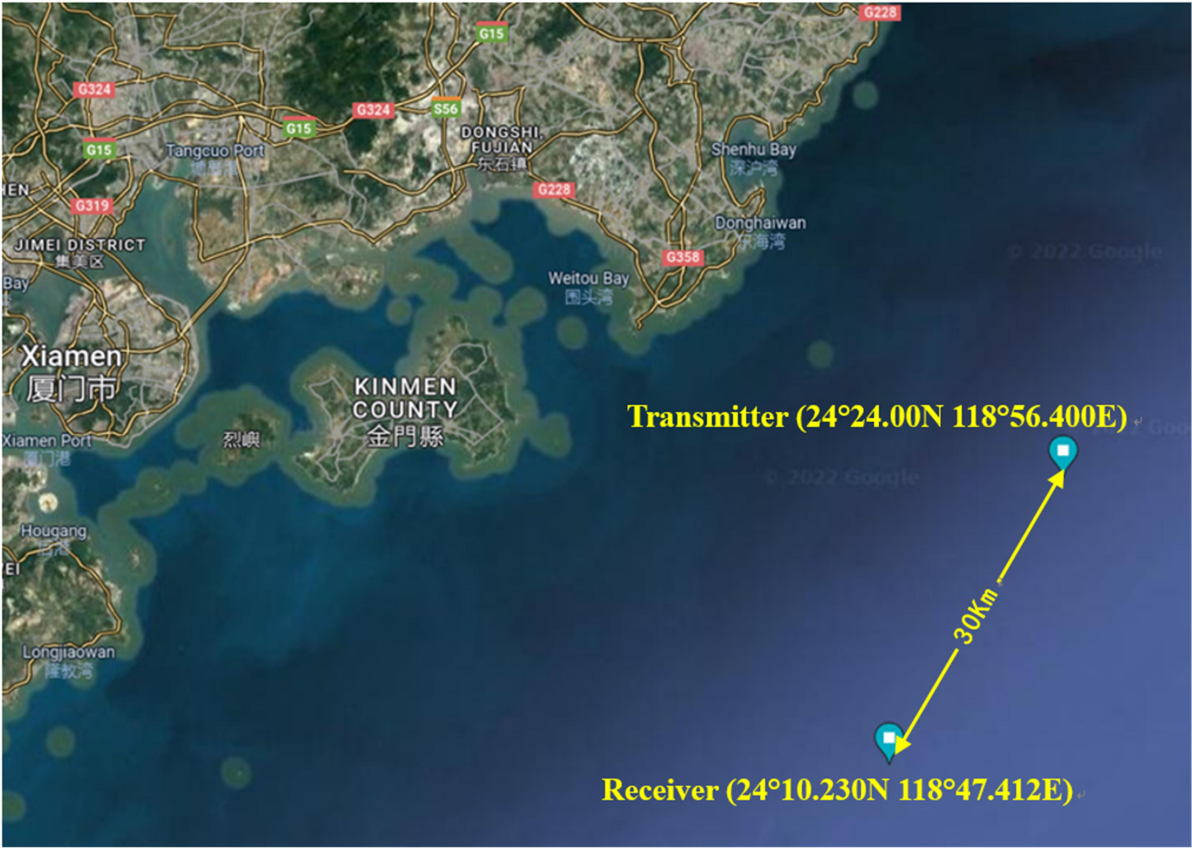}\label{fig:scenario_Strait}}
    \caption{UWA channel measurement scenario and real-world deployment.}
    \label{fig:scenario}
\end{figure}

\begin{figure*}[!t] 
    \centering
    \subfloat[UWA channel in the Furong Lake I]{\includegraphics[width=0.49\linewidth]{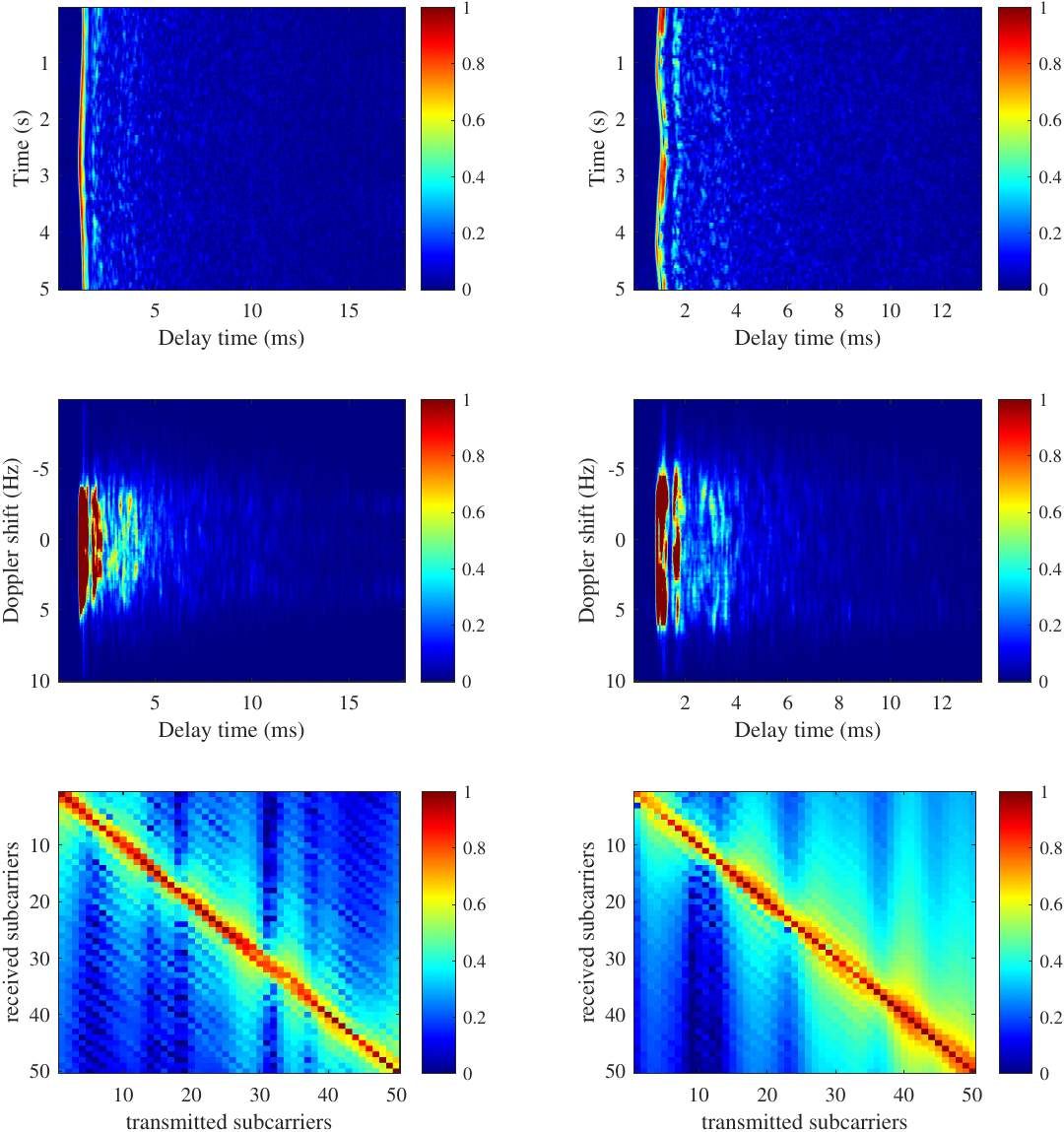}\label{fig:furongI_I}}
    \hfill
    \subfloat[UWA channel in the Furong Lake II]{\includegraphics[width=0.49\linewidth]{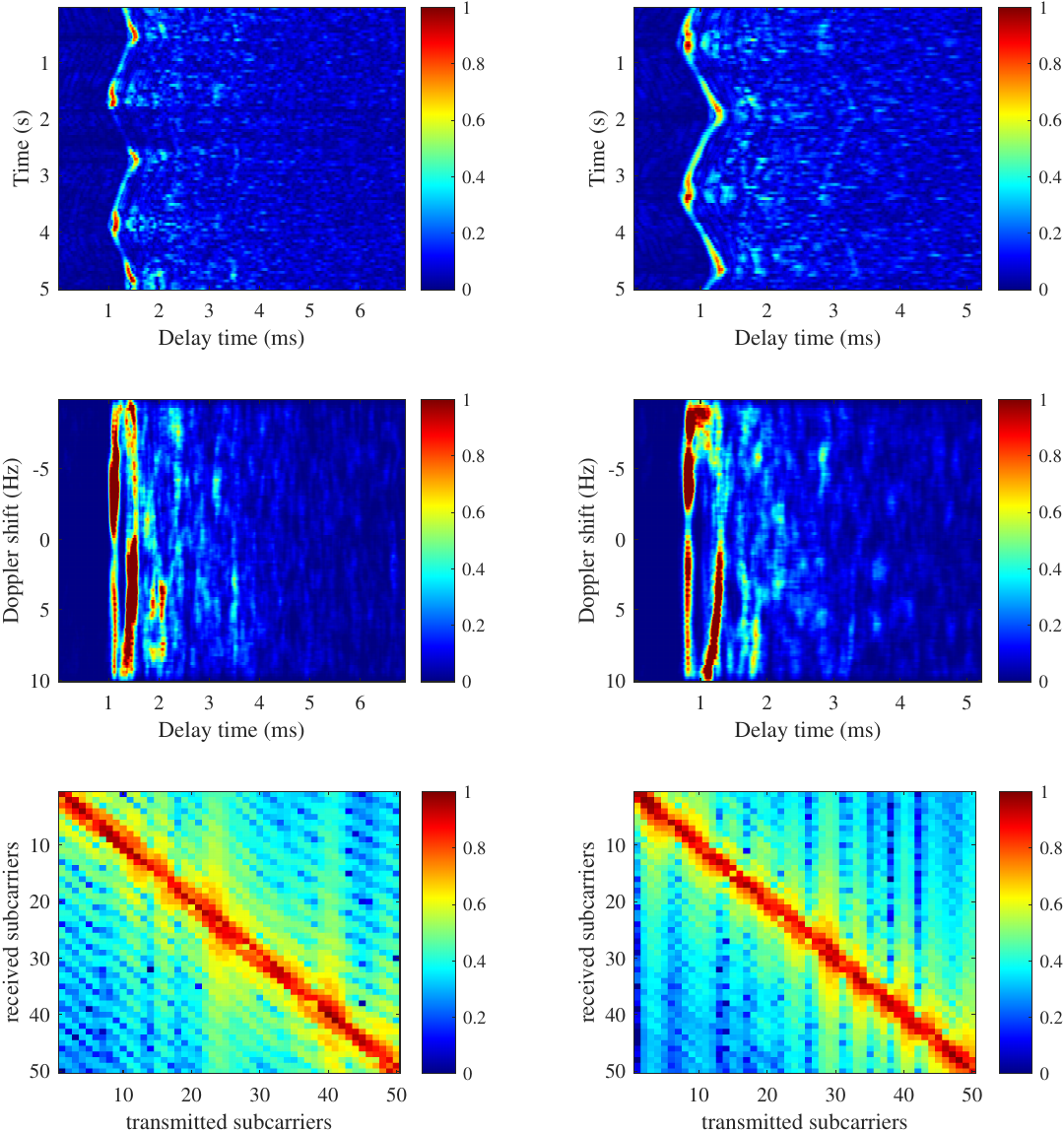}\label{fig:furongII_II}}
    \vspace{-2mm}  
    \subfloat[UWA channel in the Taiwan Strait I]{\includegraphics[width=0.49\linewidth]{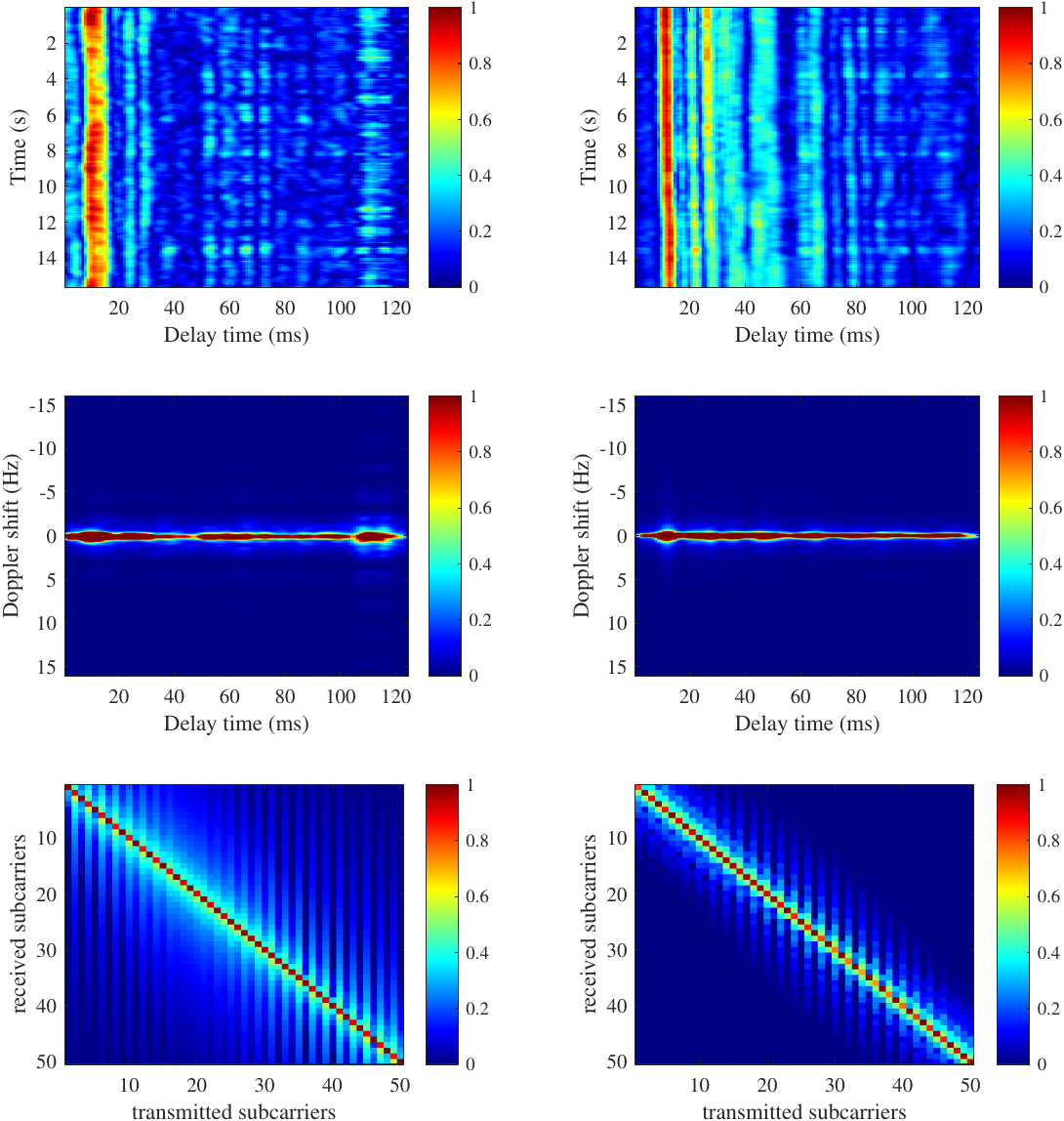}\label{fig:taiwanI_I}}
    \hfill
    \subfloat[UWA channel in the Taiwan Strait II]{\includegraphics[width=0.49\linewidth]{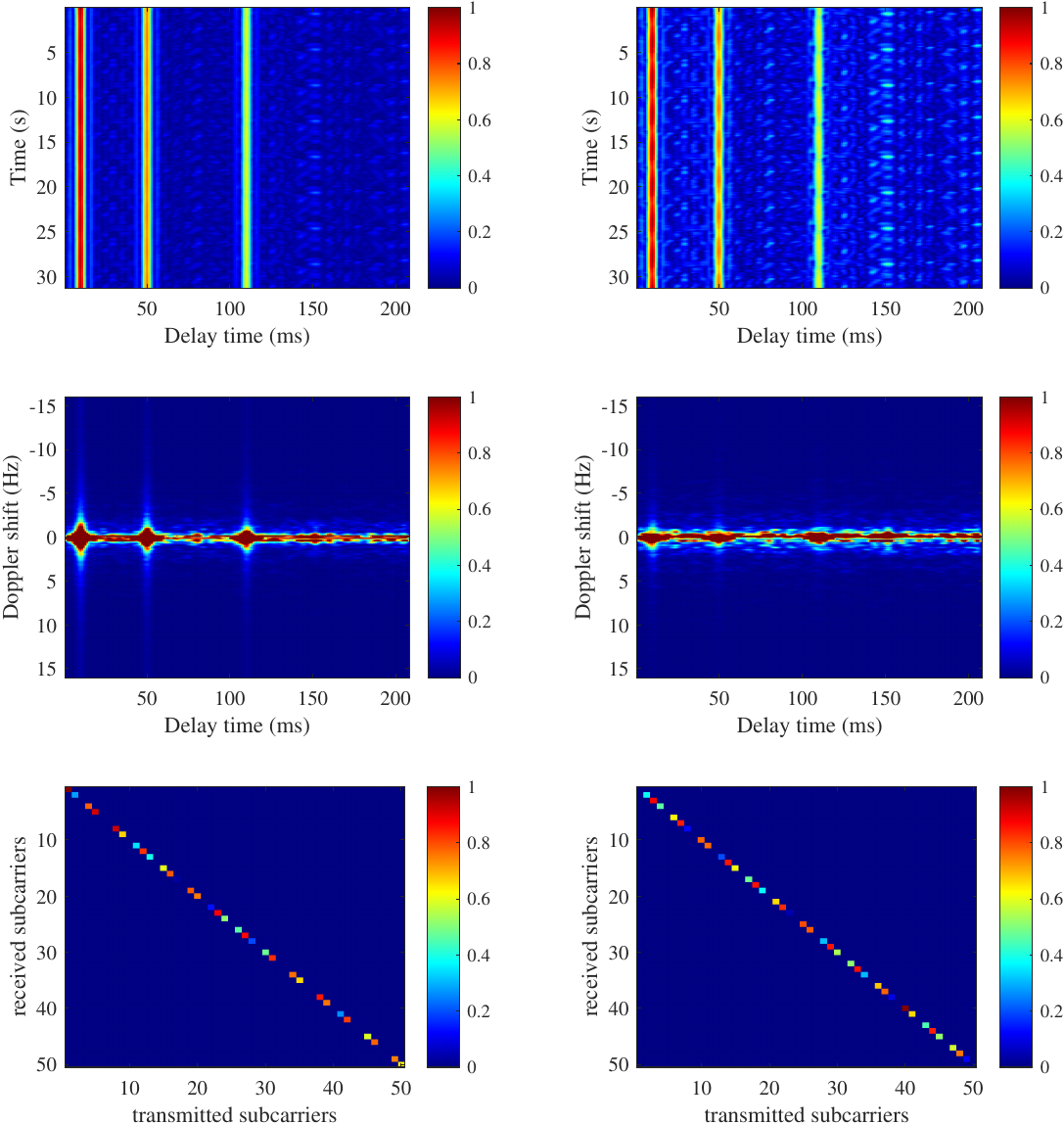}\label{fig:taiwanII_II}}
    \caption{UWA channel measurements conducted in the Furong Lake and the Taiwan Strait, where each scenario includes the CIRs, Doppler spectrum, and frequency-domain channel matrix, illustrated using 50 subcarriers as an example.}
    \label{fig:UWA channel}
\end{figure*}

In the previous simulations, the channel state information (CSI) is assumed to be perfect, whereas practical estimation inevitably incurs errors. To quantify such inaccuracies, we introduce a metric called the Channel Error Ratio (CER)~\cite{panayirci2019sparse, wang2020new, zhang2021adaptive}, given by

\begin{equation}
\text{CER}=10\log_{10}\left(\frac{\text{average power of }\mathbf{H}_{u}}{\text{average power of }\mathbf{\Theta}_{u}}\right),
\label{eq:CER}
\end{equation}
where $ \mathbf{H}_{u} $ denotes the actual frequency-domain channel matrix, and $ \mathbf{\Theta}_{u} $ represents the channel estimation error, which is interpreted as additive noise superimposed on $ \mathbf{H}_{u} $. The CER of $\mathbf{H}_A$ and $\mathbf{H}_B$ is considered identical in the same transmission condition. We evaluate the performance of proposed algorithm schemes under varying CERs to assess their robustness to channel errors, with the results after 5 outer iterations in a single relay scenario.

Fig.~\ref{fig:BER-CER} shows that channel estimation errors have a direct impact on receiver performance. ISM-LMMSE is sensitive to channel accuracy, leading to a complete detection failure when CER = 10 dB. When $\sigma_{u} = 0.1$ m/s with a BER of $10^{-3}$, the perfect baseline still outperforms ISM-LMMSE-20 by nearly 2 dB.

In contrast, IACA-FGD demonstrates remarkable robustness to imperfect channels. Under different $\sigma_{u}$, although the performance gap between IACA-FGD-10 and the ideal case remains 2--3 dB, this gap reduces to below 0.5 dB when CER = 20 dB. This resilience arises from factor graph detection, which utilizes inter-node dependencies, maintaining stable performance even under imperfect channel conditions.

\section{Experiment Results and Discussion}\label{sec:experiment}

\begin{figure*}[!t]
    \centering
    \subfloat[BER comparison in the Furong Lake I]{\includegraphics[width=0.49\linewidth]{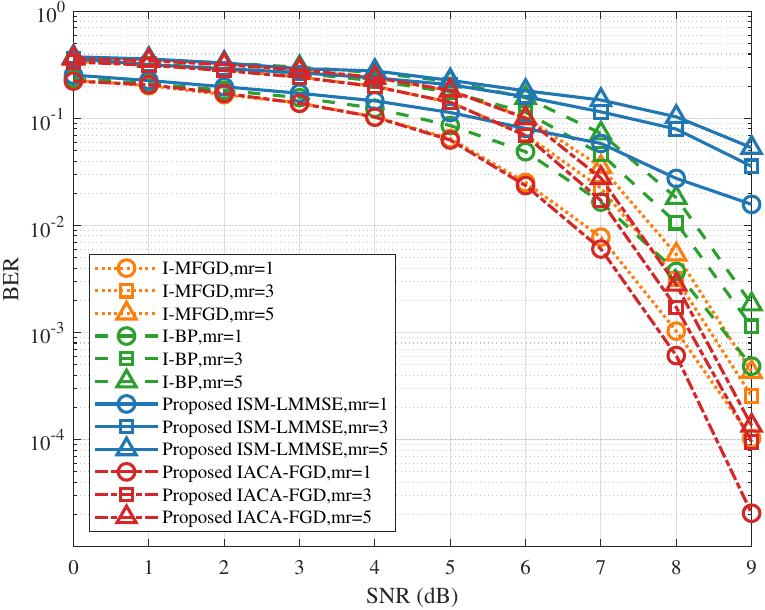}\label{fig:furongI}}
    \hfill
    \subfloat[BER comparison in the Furong Lake II]{\includegraphics[width=0.49\linewidth]{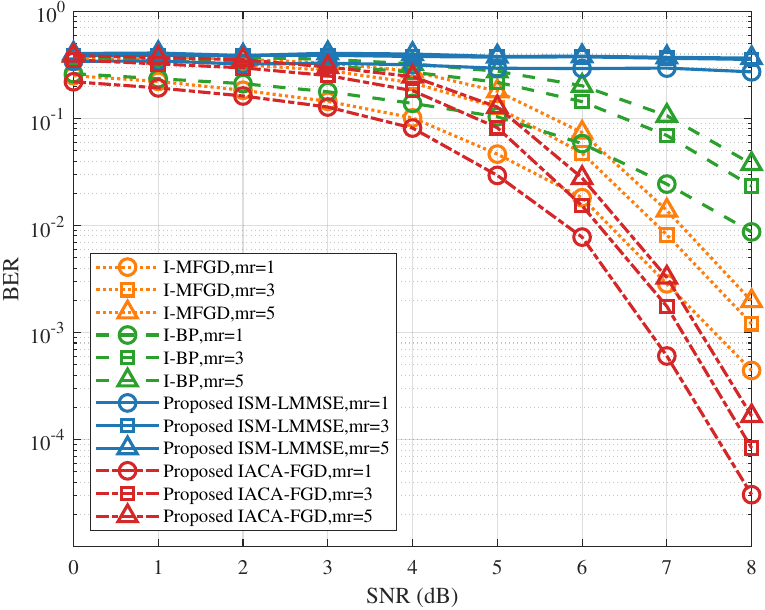}\label{fig:furongII}}
    \vspace{-1mm}  
    \subfloat[BER comparison in the Taiwan Strait I]{\includegraphics[width=0.49\linewidth]{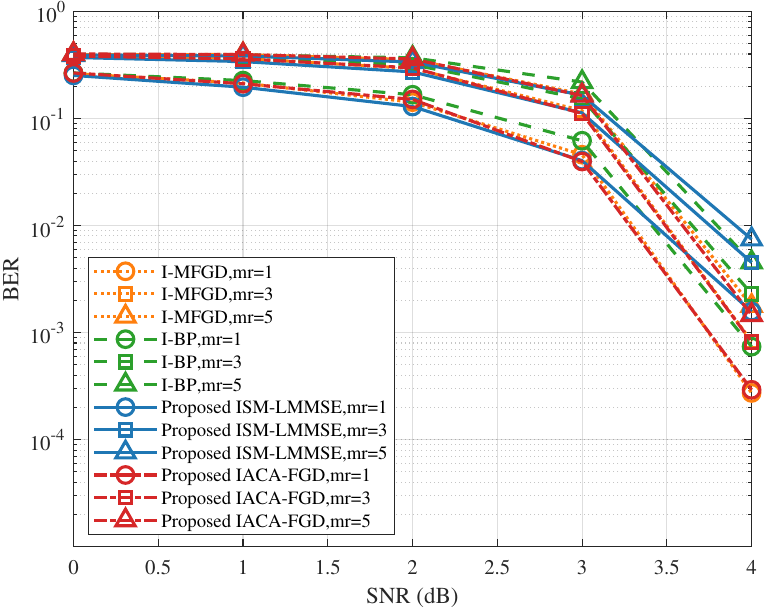}\label{fig:taiwanI}}
    \hfill
    \subfloat[BER comparison in the Taiwan Strait II]{\includegraphics[width=0.49\linewidth]{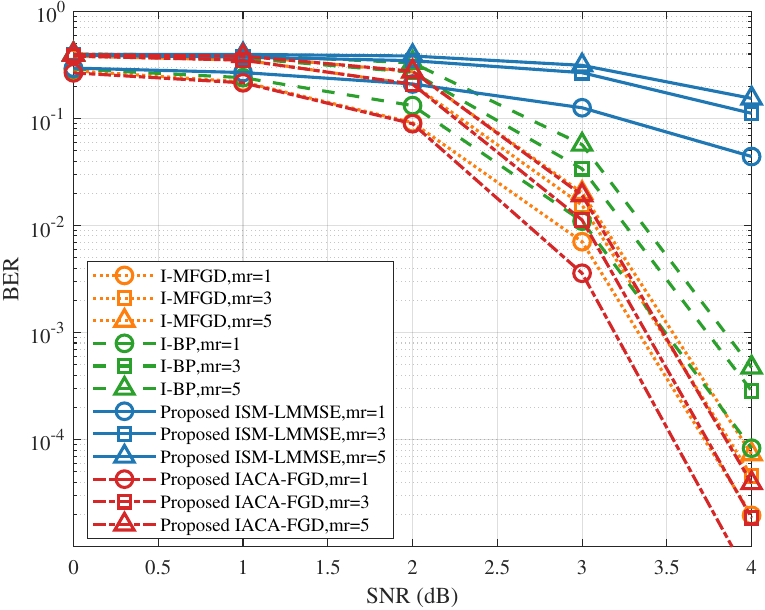}\label{fig:taiwanII}}
    \caption{Experiment performance comparison in the UWA channels measured from the Furong Lake and the Taiwan Strait.}
    \label{fig:EX}
\end{figure*}

To evaluate the performance of the proposed algorithms in practical scenarios, experiments were conducted under a single-input single-output (SISO) OFDM configuration. The bandwidth is divided into 45 subbands, with the first two subcarriers of each subband alternately using null and pilot subcarriers, ensuring perfect channel estimation. The remaining part of each subband consists of 8 data subcarriers, with the total data occupying 360 frequency positions. Information bits are encoded using binary and a rate-$ 1/3 $ irregular QC-LDPC code with BPSK modulation. we set \emph{ICI depth} in traditional schemes to 1, while the threshold for IACA-FGD is set to 90\%. Based on the performance of soft-information refinement in simulations, it is applied exclusively to IACA-FGD in the experiments, with the same parameter configuration. We set the maximum number of outer iterations to $5$, where each iteration consists of 1 detection step followed by 3 decoding iterations. Similarly, we set a multi-hop network including up to five relay nodes.

We employ the $ M $-sequence sounding method for channel estimation between the uplink and downlink nodes and the relay node. CSI is measured in the Furong Lake at Xiamen University and in the Taiwan Strait, China, as shown in Fig.~\ref{fig:scenario}. The probing signal occupies a 10~kHz bandwidth centered on a 25~kHz carrier. For the Furong Lake measurements, the time and delay sampling rates are 20~Hz and 300~kHz, respectively, with a total duration of 5~s. The Taiwan Strait experiments use sampling rates of 32~Hz and 
1~kHz, and the probing intervals are extended to 15~s and 30~s. To enable frequency-domain signal processing, we transform time-domain impulse responses into frequency-domain channel matrices by DFT.

\begin{equation}
\mathbf{H}_{u}[i+1,k+1] = 
\frac{1}{N_{F}}
\sum_{\ell=0}^{L_M-1}
\sum_{n=0}^{N_F-1}
h_{u}(n;\ell)\,
e^{j\frac{2\pi}{N_F}\left[-in + k(n-\ell)\right]},
\end{equation}
where $0 \le i, k \le N_F - 1$, $h_{u}(n;\ell)$ is the baseband CIRs in the discrete-time domain, and $L_M$ is the length of response memory.

Fig.~\ref{fig:UWA channel}\subref{fig:furongI_I} shows the quasi-static UWA channel measured in the Furong Lake, which exhibits a weak time variation, with most of the multipath delay concentrated within the first 5~ms. The mobile scenario in the same lake can be found in Fig.~\ref{fig:UWA channel}\subref{fig:furongII_II}, where the delay spread remains short, but Doppler shifts are observed. In contrast, the Taiwan Strait channels are characterized by delay. Fig.~\ref{fig:UWA channel}\subref{fig:taiwanI_I} shows typical oceanic UWA channel, featuring long multipath delays exceeding 100~ms with negligible Doppler spread while Fig.~\ref{fig:UWA channel}\subref{fig:taiwanII_II} depicts the channel with even longer delays approaching 200~ms, but the multipath variations are distributed over several delay clusters. Overall, Doppler effects are more remarkable in the lake channels, while the ocean channels exhibit longer multipath delays. When focusing on the frequency-domain channels, we observe more ICI in the lake environment, proving the Doppler shift as a key factor influencing the channel energy distribution.

Figs.~\ref{fig:EX}\subref{fig:furongI} and \ref{fig:EX}\subref{fig:furongII} illustrate the performance in the Furong Lake experiments. IACA-FGD achieves optimal performance under various channel conditions after multi-hop transmission. I-MFGD follows with a convergence performance nearly 1~dB worse than IACA-FGD. I-BP shows limited robustness, while ISM-LMMSE performs the worst, with a BER around $10^{-2}$ even at high SNRs. The performance gap among the considered schemes is more pronounced in Furong Lake II. In particular, after five relay nodes transmission, IACA-FGD achieves at least a 1~dB gain over existing schemes. Although the Furong Lake II exhibits more severe Doppler shifts, the MP-based scheme achieves better performance in this scenario than in the Furong Lake I. This improvement is attributed to the larger proportion of channel energy distributed in the sidelobe regions of Furong Lake II. In addition, the pronounced disparity between $\mathbf{H}_A$ and $\mathbf{H}_B$ in Furong Lake I further degrades the receiver performance.

Figs.~\ref{fig:EX}\subref{fig:taiwanI} and \ref{fig:EX}\subref{fig:taiwanII} show the performance results for the Taiwan Strait channels. Compared with the Furong Lake scenario, all schemes show improved performance in the Taiwan Strait. In the Taiwan Strait~I, ISM-LMMSE exhibits competitive performance after multi-hop transmission, with marginal differences compared with the other schemes. The reason is that the channel energy is highly concentrated on the dominant path. The five relays BER performance of IACA-FGD is slightly better than that of ISM-LMMSE, since the minimal sidelobe energy limits the advantages of increasing the \emph{ICI depth}. However, in Taiwan Strait II, the higher sidelobe energy leads to a decrease in ISM-LMMSE, while benefiting IACA-FGD.

\section{Conclusion}\label{sec:conclusion}

This work targets multi-hop UWA networks and introduces PNC to enhance end-to-end throughput, with a particular focus on improving relay reception reliability under MSML UWA channels. Within an iterative detection and decoding framework, three algorithms are proposed to tackle this challenge. Based on the energy distribution in the frequency channel, we propose an adaptive channel-aware factor graph detection algorithm. Considering that errors in decoding output during traditional iterative detection and decoding can degrade signal detection, we develop a soft information refinement mechanism to provide more accurate \textit{a priori} information to the detector. To offer an alternative detection with low computational complexity, we derive an superimposed-signal-based detection method. Extensive simulation and real-field experiments reveal that: (i) the proposed IACA-FGD demonstrates outstanding performance in fast time-varying channels, achieving BERs on the order of $10^{-5}$ at an SNR of 8~dB; and (ii) ISM-LMMSE serves as a low-complexity alternative in slow-varying channels, attaining BERs on the order of $10^{-4}$ at the same SNR. These results confirm the feasibility of applying PNC to real-field multi-hop UWA networks, providing a practical solution for boosting end-to-end throughput. Future work will integrate channel estimation into the iterative processing chain, enabling the development of a complete and deployable PNC relay receiver.

\ifCLASSOPTIONcaptionsoff
  \newpage
\fi

\bibliographystyle{IEEEtran}
\bibliography{refs}

\input{bio}



\end{document}

%% file: bio.tex
\begin{IEEEbiography}{Gewei Zhang} received the BE degree in communication engineering from Huaqiao University, Xiamen, China, in 2024. He is currently working toward the ME degree in communication engineering with the Department of Information and Communication Engineering, Xiamen University, Xiamen, China. His research interests include underwater acoustic communication and networks and signal processing.
\end{IEEEbiography}

\vspace{-0.5in}

\begin{IEEEbiography}{Deqing Wang}(Member, IEEE) received his Ph.D. degree in Communication Engineering from Xiamen University, China, in 2013. He is currently an Associate Professor at the Key Laboratory of Underwater Acoustic Communication and Marine Information Technology, Xiamen University, Ministry of Education, Xiamen, China. He was a visiting scholar at Georgia Institute of Technology from 2016 to 2017. His research interests lie in underwater acoustic communication and networking, signal processing, with a particular focus on physical-layer network coding (PNC).
\end{IEEEbiography}

\vspace{-0.5in}

\begin{IEEEbiography}{Lizhao You}(Member, IEEE) received the Ph.D. degree from The Chinese University of Hong Kong, China, in 2016, and the B.S. and M.E. degrees from Nanjing University, China, in 2009 and 2013, respectively. He joined Huawei Technologies Co., Ltd. after his graduation, and worked for four years. He joined Xiamen University, China, in 2021, and is currently an Assistant Professor in the School of Informatics. His research interests include wireless communication and networks, and computer networks.
\end{IEEEbiography}

\vspace{-0.5in}

\begin{IEEEbiography}{Xiangming Cai}(Member, IEEE) received the Ph.D. degree in communication and information systems from Xiamen University, Xiamen, China, in December 2021. From March 2022 to 2023, he was a Research Fellow with the Engineering Product Development Pillar, Singapore University of Technology and Design, Singapore. From April 2023 to 2024, he was a Post-Doctoral Fellow with the Department of Electrical and Electronic Engineering, The Hong Kong Polytechnic University, Hong Kong. Since May 2024, he has been an Associate Professor with the School of Physics and Information Engineering, Fuzhou University, Fuzhou, China. His research interests include wireless communications and underwater acoustic communications. Dr. Cai was a recipient of the Best Paper Award of the 2021 IEEE Asia–Pacific Conference on Communications. He was also a recipient of an Exemplary Reviewer of IEEE TRANSACTIONS ON COMMUNICATIONS.
\end{IEEEbiography}

\vspace{-0.5in}

\begin{IEEEbiography}{Liqun Fu}(S'08-M'11-SM'17) is a Full Professor of the School of Informatics at Xiamen University, China. She received her Ph.D. Degree in Information Engineering from The Chinese University of Hong Kong in 2010. Her research interests are mainly in communication theory, optimization theory, game theory, and learning theory, with applications in wireless networks. She is on the editorial board of IEEE Transactions on Mobile Computing (TMC), IEEE Communications Letters (CL) and the Journal of Communications and Information Networks (JCIN). She served as the Technical Program Co-Chair of IEEE/CIC ICCC 2021 and the GCCCN Workshop of the IEEE INFOCOM 2014, the Publicity Co-Chair of the GSNC Workshop of the IEEE INFOCOM 2016, and the Web Chair of the IEEE WiOpt 2018. She also serves as a TPC member for many leading conferences in communications and networking, such as the IEEE INFOCOM, ICC, and GLOBECOM.
\end{IEEEbiography}